\documentclass[11pt,english,preprint,flushrt]{aastex}
\usepackage{amsmath}
\usepackage{graphics}
\usepackage{amssymb}
\usepackage{subfigure}
\usepackage{natbib}

\shortauthors{Geroyannis and Papasotiriou}
\shorttitle{The Turn-Over Scenario, II}

\begin{document}

\title{A TURN-OVER SCENARIO FOR ROTATING MAGNETIC \\
       WHITE DWARFS: MODELS WITH SEVERAL VALUES \\
       OF MASS, ANGULAR MOMENTUM, \\
       AND MAGNETIC FIELD}

\author{V. S. Geroyannis and P. J. Papasotiriou}

\affil{Astronomy Laboratory, Department of Physics, University of Patras,
Greece,}

\affil{GR-26500 PATRAS, GREECE}

\email{vgeroyan@physics.upatras.gr, papasot@physics.upatras.gr}

\begin{abstract}
We study a white dwarf model with differential rotation and magnetic
field, for which the symmetry axis of the toroidal field, the magnetic
axis of the poloidal field, and the principal axis \( I_{3} \) coincide
permanently; the common axis defined this way is called {}``magnetic
symmetry axis''. Furthermore, the magnetic symmetry axis inclines
at a small angle \( \chi  \) relative to the spin axis of the model;
this angle is called {}``obliquity angle'' or {}``turn-over angle''.
Such a model is almost axisymmetric and undergoes an early evolutionary
phase of secular timescale, characterized by the fact that the moment
of inertia along the spin axis, \( I_{zz}\simeq I_{33} \), is greater
than the moments of inertia along the (almost) equatorial axes, \( I_{11}=I_{22} \),
since rotation and poloidal field (both responsible for the oblateness
of the model) dominate over the toroidal field (responsible, in turn,
for the prolateness of the model). During this early evolutionary
phase, the model suffers from secular angular momentum loss due to
weak magnetic dipole radiation activated by the poloidal field. Such
an angular momentum loss leads gradually to a situation of dynamical
asymmetry with \( I_{11}>I_{33} \). However, dynamically asymmetric
configurations tend to turn over spontaneously, that is, to rotate
about axis with moment of inertia greater than \( I_{33} \) with
angular momentum remaining invariant. So, the fate of a dynamically
asymmetric configuration is to become an oblique rotator and, eventually,
a perpendicular rotator. During the so-called {}``turn-over phase'',
the turn over angle, \( \chi  \), increases spontaneously up to \( \sim 90^{\circ } \)
on a {}``turn-over timescale'', \( t_{\mathrm{TOV}} \), since the
rotational kinetic energy of the model decreases from a higher level
when \( \chi \simeq 0^{\circ } \) (aligned rotator) to a lower level
when \( \chi \simeq 90^{\circ } \) (perpendicular rotator). At this
lower level the model reaches the state of least energy consistent
with its prescribed angular momentum and magnetic field. The excess
rotational kinetic energy due to differential rotation is totally
dissipated due to the action of turbulent viscosity in the convective
regions of the model. In the present paper, we study numerically the
so-called {}``turn-over scenario'' (i.e., an evolutionary scenario,
which takes into account the turn-over phase) for white dwarf models
of several masses, angular momenta, and magnetic fields.
\end{abstract}


\section{Introduction}

In a recent paper \citep[hereafter Paper I, and references therein]{ger01},
the turn-over scenario has been studied numerically for rotating magnetic white dwarfs \citep{ger02}. The turn-over scenario, hereafter TOV scenario, deals with the problem of rotational evolution of a star by taking into account the gradual increase of the turn-over angle, i.e., the angle between the magnetic symmetry axis and the spin axis of the star.

In the TOV scenario, we assume that the differentially rotating white
dwarf model is initially almost axisymmetric (i.e., its turn-over
angle \( \chi  \) is small: \( \chi \leq 2^{\circ } \), say) and
undergoes an {}``early evolutionary phase'', during which rotation
and poloidal field prevail against the toroidal field, yielding oblate
configurations with \( I_{33}>I_{11} \) (where \( I_{33} \) is the
moment of inertia along the magnetic symmetry axis, almost coinciding
with the spin axis, and \( I_{11}=I_{22} \) are the moments of inertia
along the other two principal axes).

However, due to a weak magnetic dipole radiation activated by the
poloidal field, the toroidal field becomes gradually more effective,
and eventually leads the model to the so-called {}``late evolutionary
phase'', during which the model suffers from {}``dynamical asymmetry'':
\( I_{11}>I_{33} \).

Dynamical asymmetry leads the model to the so-called {}``turn-over
phase'', during which the turn-over angle, \( \chi  \), increases
spontaneously up to \( \sim 90^{\circ } \) on a turn-over timescale,
\( t_{\mathrm{TOV}} \). The terminal model rotates about its \( I_{1} \)
axis, coinciding with the invariant angular momentum axis, and occupies
the state of least energy consistent with its angular momentum and
magnetic field. The excess energy due to differential rotation, defined
by the angular velocity component \( \Omega _{3} \) along the spontaneously
turning over \( I_{3} \) axis (permanently coinciding with the magnetic
symmetry axis), is dissipated down to zero due to the efficient action
of turbulent viscosity in the convective zone of the model. So, the
terminal model does not rotate about its magnetic symmetry axis. Furthermore,
it seems difficult for the terminal model to sustain differential
rotation along its \( I_{1} \) axis, mainly due to the destructive
action of the poloidal field. In particular, there is a competition
between the efforts of the magnetic stresses to remove rotational
nonuniformities, and those of the rotational velocities to bury and
destroy the magnetic flux. If the magnetic field and the electrical
conductivity have appropriate values (see especially \S~7 of Paper I),
then the magnetic field prevails and removes all the nonuniformities
of rotation. So, the terminal model rotates rigidly about its \( I_{1} \)
principal axis with angular velocity \( \Omega _{1} \).

In Paper I, the aim was to compute the so-called {}``optimal
values'' for the angular momentum, \( L_{xx} \), the average surface
poloidal field, \( B_{s} \), and the time evolution parameter, \( \delta  \),
under which the model starts its turn-over phase (Paper I, \S~6).
In the present paper, on the other hand, our aim is slightly different.
In particular, we shall study several possible turn-over scenarios,
corresponding to several indicative values \( L_{xx} \) and \( B_{s} \).
We shall describe in detail our computations in the following sections.

\section{The model, the numerical method, and the computations}

We shall use hereafter definitions and symbols identical to those
in Paper I.

The model and the numerical method used in the study of the TOV scenario
have been thoroughly discussed in Paper I. All the technical details of that paper are closely related
to each other in a way making difficult to summarize some issues without
summarizing the remaining issues of the whole matter. Since our intention
is to avoid repeating that paper here, we clearly consider the present
paper as continuation of Paper I.

For the computations of the present paper, we assume that the angular
momentum, \( L_{xx} \), and the average surface poloidal field, \( B_{s} \),
of the starting model are free model parameters (in Paper I,
instead, we have computed optimal values for these quantities;
see especially \S~6 of that paper). Accordingly, the spin-down time
rate due to turn-over, \( \overset {\cdot }{P}_{\mathrm{TOV}} \),
becomes now a dependent model parameter taking several values for
several choices of the whole set of the free model parameters.

We study the TOV scenario for three {}``reference'' cases of white
dwarfs; namely, (a) \( M=M_{\mathrm{REF}1}=0.6\, M_{\odot } \) and
\( P=P_{\mathrm{REF}1}=142\, \mathrm{s} \), (b) \( M=M_{\mathrm{REF}2}=0.9\, M_{\odot } \)
and \( P=P_{\mathrm{REF}2}=33\, \mathrm{s} \), (c) \( M=M_{\mathrm{REF}3}=1.32\, M_{\odot } \)
and \( P=P_{\mathrm{REF}3}=8\, \mathrm{s} \). For all these cases,
we assume a {}``reference'' period time rate, \( |\overset {\cdot }{P}_{\mathrm{REF}}|=5\times 10^{-13}\, \mathrm{s}\, \mathrm{s}^{-1} \),
which should be dominant in the absence of the spin-down time rate
due to turn-over, \( \overset {\cdot }{P}_{\mathrm{TOV}} \). Our
purpose is to study the variation of the ratio \( \overset {\cdot }{P}_{\mathrm{TOV}}/|\overset {\cdot }{P}_{\mathrm{REF}}| \)
with the free parameters \( L_{xx} \) and \( B_{s} \), and to find
when \( \overset {\cdot }{P}_{\mathrm{TOV}} \) becomes of comparable
or even higher significance with respect to \( \overset {\cdot }{P}_{\mathrm{REF}} \)
(the latter could be due to accretional activity around the star and
its actual sign --- spin-down for plus, spin-up for minus --- is insignificant
here). All the starting models are assumed to be in a state of {}``physical
differential rotation'', \( F_{r}=1 \).

For each case, we consider seven {}``representative'' values for
(a) the angular momentum, \( L_{xx} \), and (b) the average surface
poloidal field, \( B_{s} \). The lower value of angular momentum
adopted corresponds to a turn-over phase just started, while the higher
angular momentum adopted corresponds to a turn-over phase near its
end.

\begin{figure*}
{\centering \resizebox*{1\textwidth}{!}{\rotatebox{-90}{\includegraphics{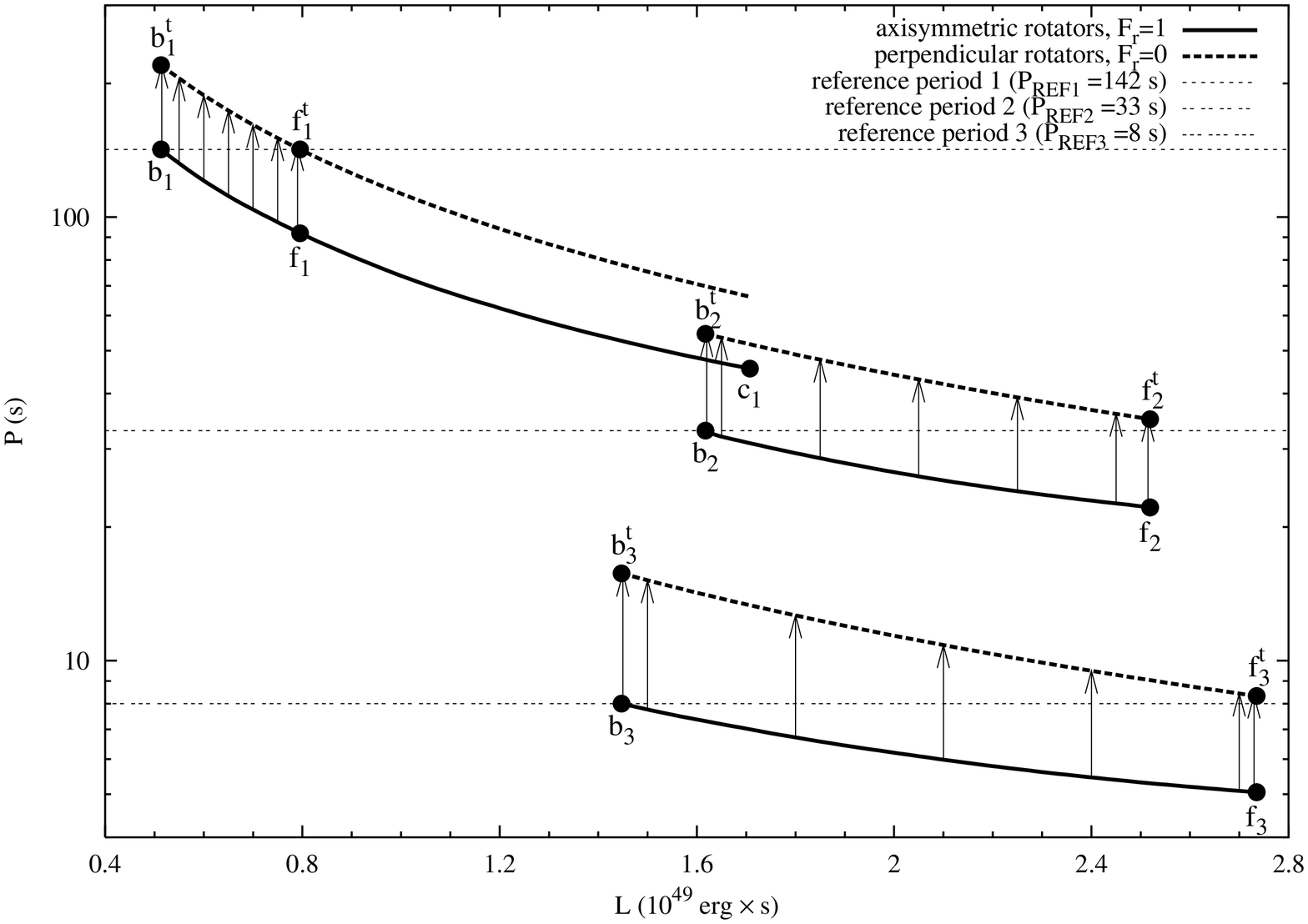}}} \par}

\caption{Plot of period at center, \protect\( P\protect \), as a function
of angular momentum, \protect\( L\protect \), for both the aligned
rotators (thick solid curves) and the perpendicular rotators (thick
dashed curves). Points \protect\( b_{1},b_{2},b_{3}\protect \): models
with central periods equal to the {}``reference'' periods studied
in this paper, \protect\( P_{\mathrm{REF}1}=142\, \mathrm{s},P_{\mathrm{REF}2}=33\, \mathrm{s},P_{\mathrm{REF}3}=8\, \mathrm{s}\protect \);
points \protect\( b_{1}^{t},b_{2}^{t},b_{3}^{t}\protect \): terminal
models corresponding to the starting models \protect\( b_{1},b_{2},b_{3}\protect \);
points \protect\( f_{1},f_{2},f_{3}\protect \): starting models with
maximum possible angular momentum; points \protect\( f_{1}^{t},f_{2}^{t},f_{3}^{t}\protect \):
terminal models corresponding to the starting models \protect\( f_{1},f_{2},f_{3}\protect \);
point \protect\( c_{1}\protect \): model for which \protect\( I_{33}=I_{11}=I_{22}\protect \)
(note that corresponding models for the second and third case, \protect\( c_{2},c_{3}\protect \),
coincide with the points \protect\( f_{2},f_{3}\protect \), respectively).
Vertical arrows show possible turn-over scenarios currently in progress,
which are studied numerically in this paper.\label{fig:P(L)}}
\end{figure*}

The turn-over phase can be easily understood in a graph showing the
period as a function of angular momentum, \( P\left( L\right)  \),
for both the aligned and perpendicular rotators (Fig.~\ref{fig:P(L)}).
The angular momentum remains constant during the turn-over. Consequently,
a turn-over phase is shown in the \( P\left( L\right)  \) graph as
a vertical arrow, connecting the starting model (located at the aligned
rotator's curve) with the terminal model (located, in turn, at the
perpendicular rotator's curve).

\section{Results and discussion}

Computed parameters for the aligned and the perpendicular rotators
are given in Tables~\ref{tab:060_Aligned_constant_for_Bs}--\ref{tab:132_Perpendicular_variable_for_Bs}.
In Tables~\ref{tab:060_Aligned_constant_for_Bs}, \ref{tab:090_Aligned_constant_for_Bs},
and \ref{tab:132_Aligned_constant_for_Bs}, we give parameters regarding
the aligned rotators. These parameters are almost independent of the
average surface poloidal field, \( B_{s} \); namely, the central
period, \( P_{xx} \), the rotational kinetic energy, \( T_{xx} \),
the moments of inertia \( I_{11} \) and \( I_{33} \) along the principal
axes \( I_{1} \) and \( I_{3} \), the average surface toroidal field,
\( \left\langle H_{ts}\right\rangle  \), the maximum toroidal field,
\( H_{t\left[ \mathrm{max}\right] } \), the average toroidal field,
\( \left\langle H_{t}\right\rangle  \), and the ratio \( \omega ^{-1}_{\mathrm{e}}=\Omega _{\mathrm{c}}/\Omega _{\mathrm{e}} \),
where \( \Omega _{\mathrm{c}} \), \( \Omega _{\mathrm{e}} \) are
the central and equatorial angular velocities, respectively. The corresponding
parameters for the perpendicular rotators are given in Tables~\ref{tab:060_Perpendicular_constant_for_Bs},
\ref{tab:090_Perpendicular_constant_for_Bs}, and \ref{tab:132_Perpendicular_constant_for_Bs}.
Here, \( P_{\mathrm{RR}} \) is the central period and \( T_{\mathrm{RR}} \)
is the rotational kinetic energy of the perpendicular rotator.

Parameters regarding the aligned rotators and varying with the average
surface poloidal field, \( B_{s} \), are given in Tables~\ref{tab:060_Aligned_variable_for_Bs},
\ref{tab:090_Aligned_variable_for_Bs}, and \ref{tab:132_Aligned_variable_for_Bs};
namely, the corresponding poloidal magnetic parameter, \( \beta ^{p}_{*} \),
the surface Alfv\'en speed, \( V_{As} \), the surface Alfv\'en time,
\( t_{As} \), the maximum poloidal field, \( H_{p\left[ \mathrm{max}\right] } \),
and the average poloidal field, \( \left\langle H_{p}\right\rangle  \).
The corresponding parameters for the perpendicular rotators are given
in Tables~\ref{tab:060_Perpendicular_variable_for_Bs}, \ref{tab:090_Perpendicular_variable_for_Bs},
and \ref{tab:132_Perpendicular_variable_for_Bs}.

General results concerning the turn-over phase are given in Tables~\ref{tab:060_Turn_Over},
\ref{tab:090_Turn_Over}, and \ref{tab:132_Turn_Over}. Parameters tabulated
in these tables are the magnetic flux, \( f \), the turn-over timescale,
\( t_{\mathrm{TOV}} \), the spin-down time rate due to turn-over,
\( \overset {\cdot }{P}_{\mathrm{TOV}} \), the turn-over timescale
in units of the starting central period, \( N_{xx}=t_{\mathrm{TOV}}/P_{xx} \),
the present turn-over time, \( t_{\mathrm{now}} \), and the power
loss due to turn-over, \( \overset {\cdot }{T} \). These tables also
contain three parameters which are independent of the average surface
poloidal field; namely, the rigid rotation amplification ratio, \( A_{r} \),
the current turn-over angle, \( \chi _{\mathrm{now}} \), and the
current differential rotation strength, \( F_{r\left[ \mathrm{now}\right] } \).

Our numerical results reveal that the turn-over timescale, \( t_{\mathrm{TOV}} \),
varies from \( \sim 0.4 \) to \( \sim 7300 \) million years, dependent
on the mass of the model, the angular momentum, and the average surface
poloidal field of the starting model. The turn-over angles \( \chi _{\mathrm{now}} \),
corresponding to the present turn-over times \( t_{\mathrm{now}} \),
vary from \( \sim 3^{\circ } \) to \( \sim 90^{\circ } \).

\begin{figure*}
{\centering \subfigure[3D plot]{\resizebox*{0.49\textwidth}{!}{\rotatebox{-90}{\includegraphics{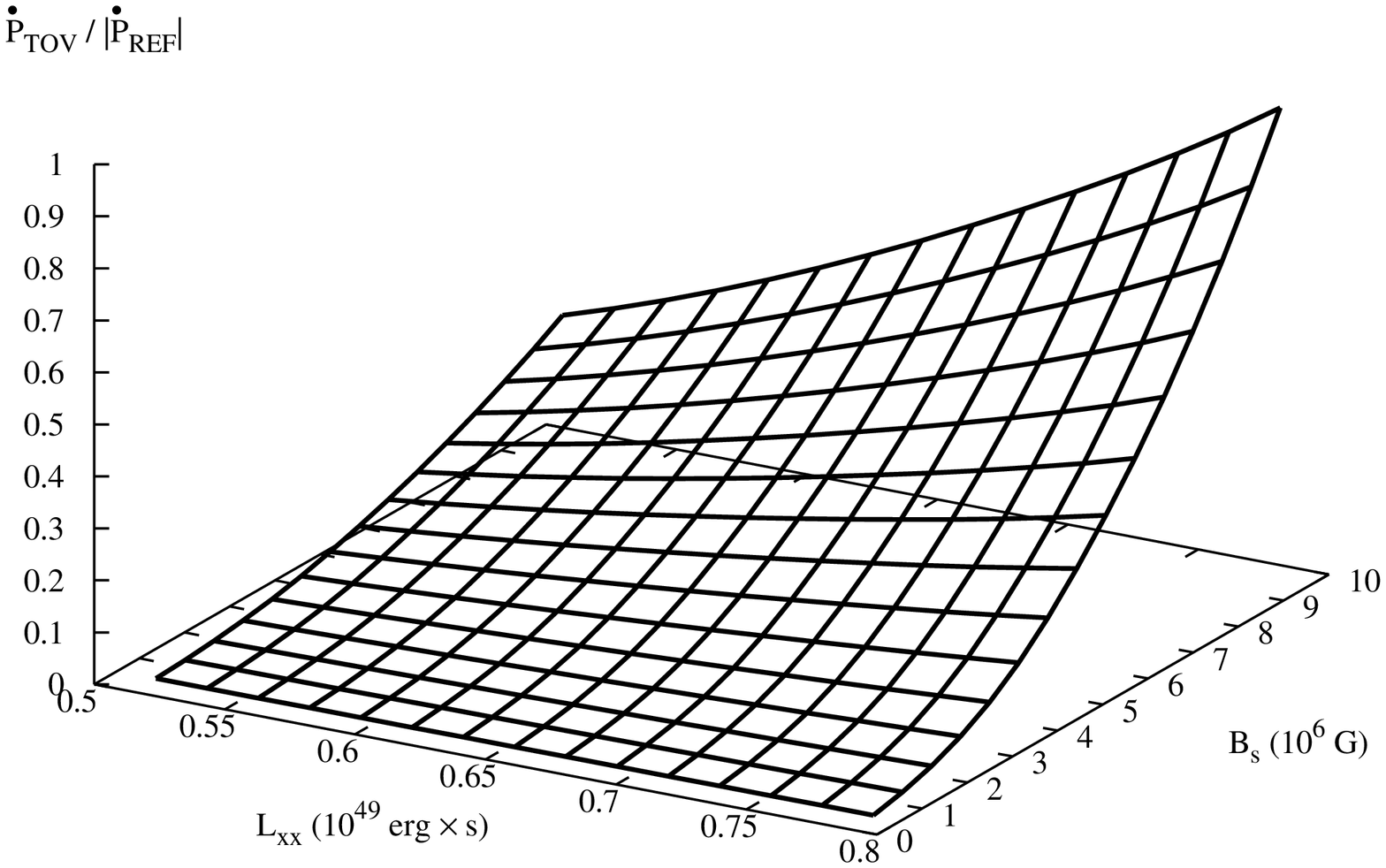}}}} \subfigure[Contour plot]{\resizebox*{0.49\textwidth}{!}{\rotatebox{-90}{\includegraphics{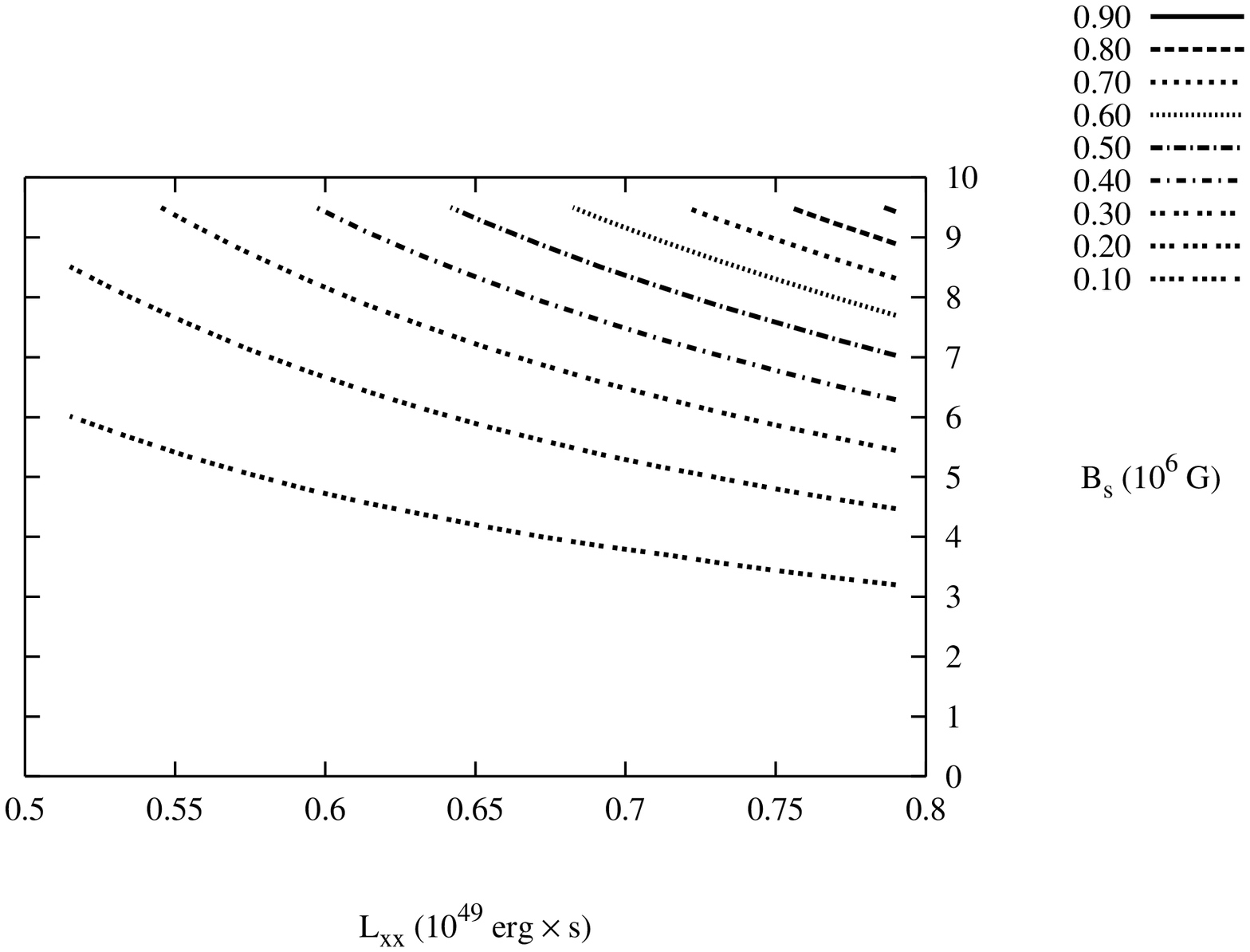}}}} \par}

\caption{The ratio \protect\( \left( dP_{\mathrm{TOV}1}/dt\right) /\left| dP_{\mathrm{REF}1}/dt\right| \protect \)
as a function of \protect\( L_{xx}\protect \) and \protect\( B_{s}\protect \)
for the case \protect\( M=0.6\, M_{\odot }\protect \).\label{fig:060_P_timerate}}
\end{figure*}

\begin{figure*}
{\centering \subfigure[3D plot]{\resizebox*{0.49\textwidth}{!}{\rotatebox{-90}{\includegraphics{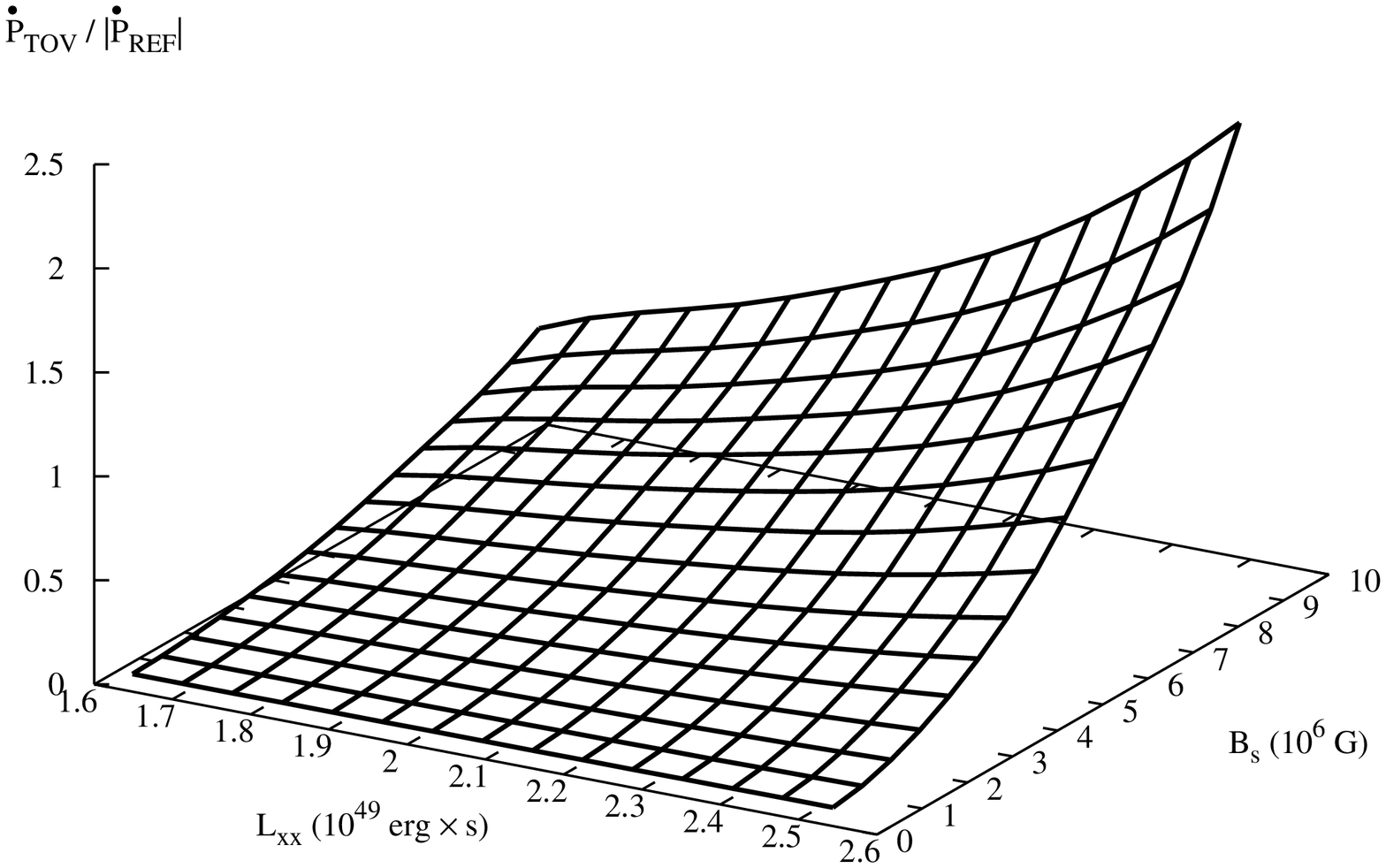}}}} \subfigure[Contour plot]{\resizebox*{0.49\textwidth}{!}{\rotatebox{-90}{\includegraphics{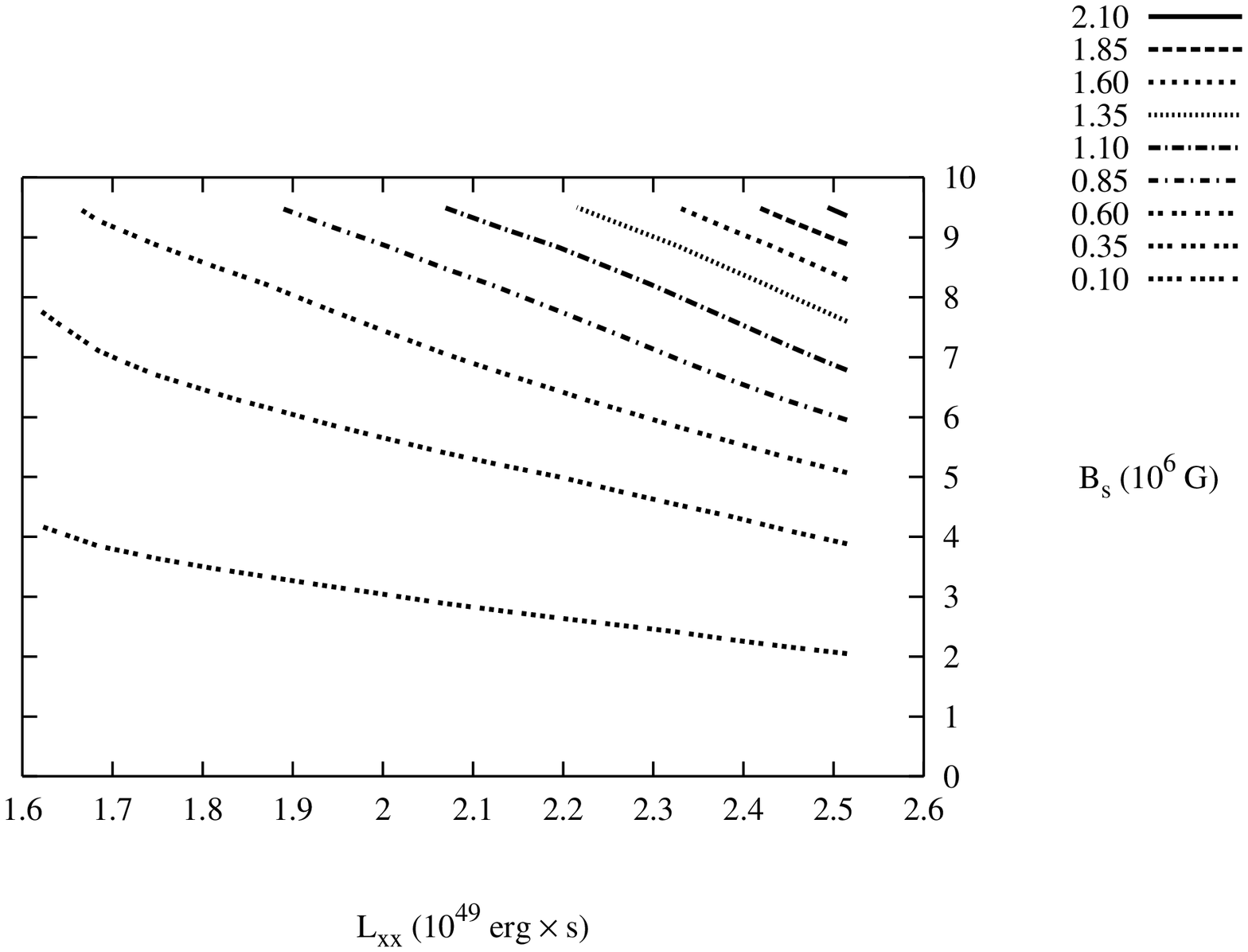}}}} \par}

\caption{The ratio \protect\( \left( dP_{\mathrm{TOV}2}/dt\right) /\left| dP_{\mathrm{REF}2}/dt\right| \protect \)
as a function of \protect\( L_{xx}\protect \) and \protect\( B_{s}\protect \)
for the case \protect\( M=0.9\, M_{\odot }\protect \).\label{fig:090_P_timerate}}
\end{figure*}

\begin{figure*}
{\centering \subfigure[3D plot]{\resizebox*{0.49\textwidth}{!}{\rotatebox{-90}{\includegraphics{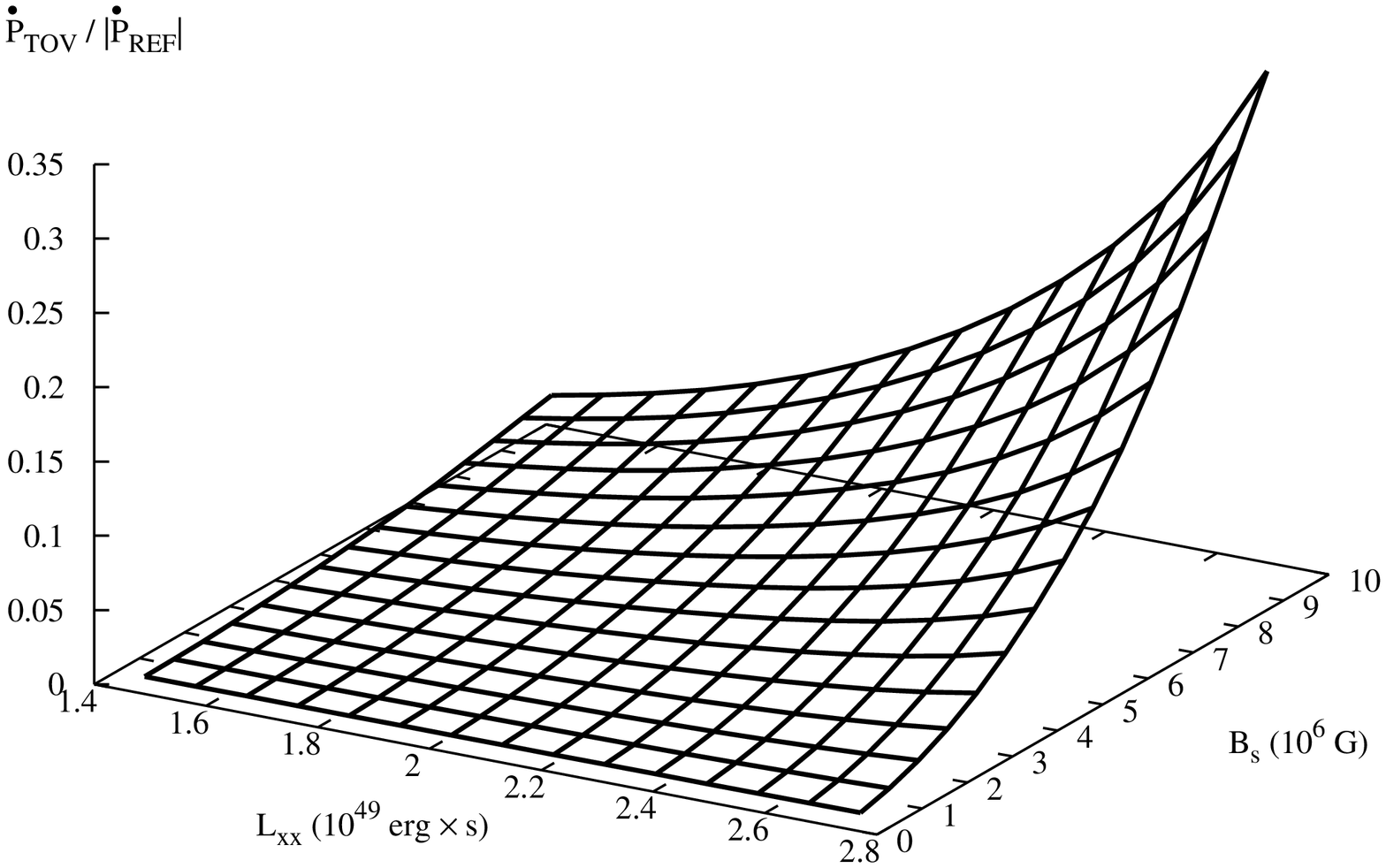}}}} \subfigure[Contour plot]{\resizebox*{0.49\textwidth}{!}{\rotatebox{-90}{\includegraphics{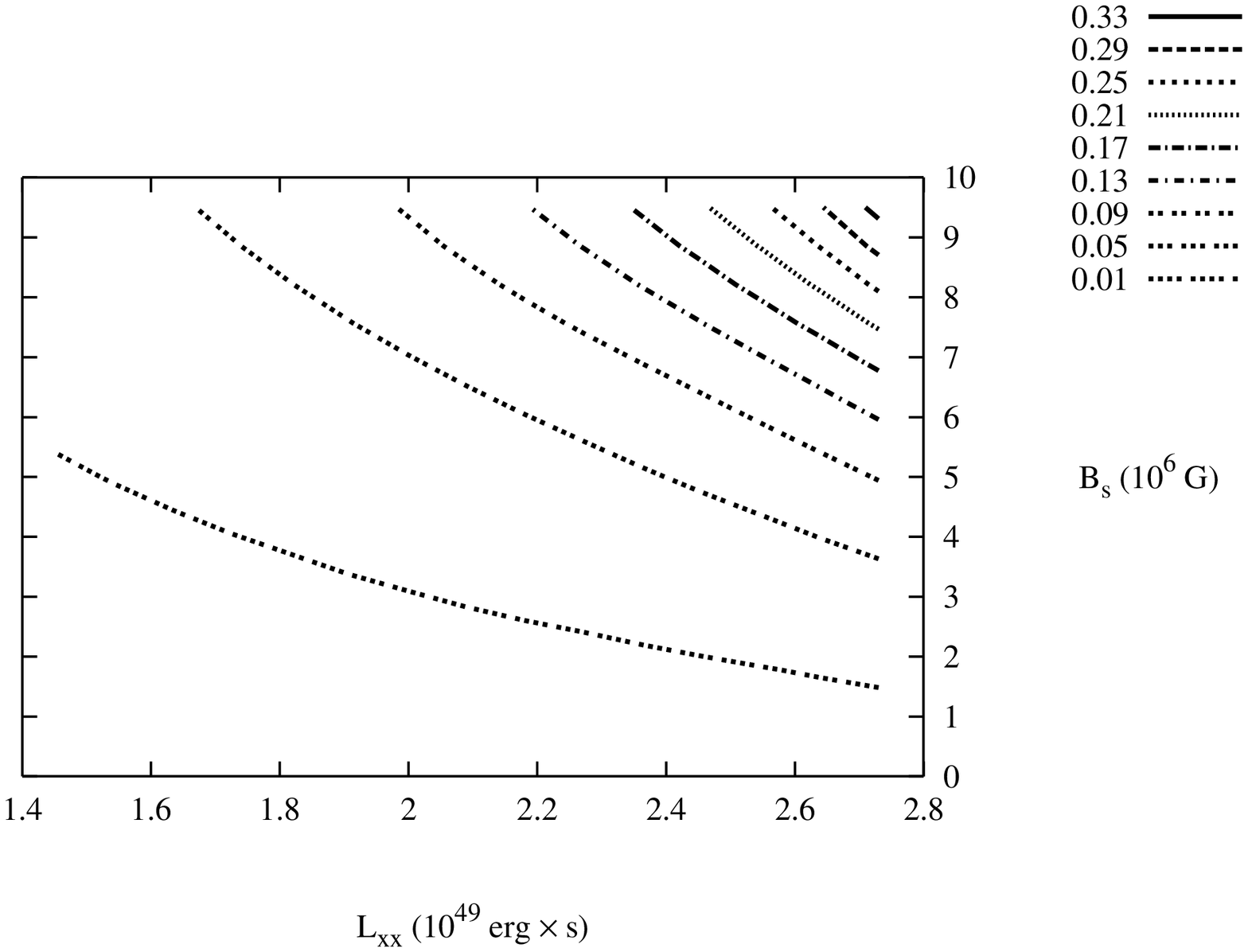}}}} \par}

\caption{The ratio \protect\( \left( dP_{\mathrm{TOV}3}/dt\right) /\left| dP_{\mathrm{REF}3}/dt\right| \protect \)
as a function of \protect\( L_{xx}\protect \) and \protect\( B_{s}\protect \)
for the case \protect\( M=1.32\, M_{\odot }\protect \).\label{fig:132_P_timerate}}
\end{figure*}

An issue of particular interest is the estimated values for the spin-down
time rate due to turn-over, \( \overset {\cdot }{P}_{\mathrm{TOV}} \),
which vary from \( 4.3\times 10^{-17}\, \mathrm{s}\, \mathrm{s}^{-1} \)
to \( 1.1\times 10^{-12}\, \mathrm{s}\, \mathrm{s}^{-1} \). From
columns 1 and 4 of Tables~\ref{tab:060_Turn_Over}, \ref{tab:090_Turn_Over},
and \ref{tab:132_Turn_Over}, it is apparent that, for low
values of the surface poloidal field, the turn-over spin-down is negligible,
since \( \overset {\cdot }{P}_{\mathrm{TOV}}\ll |\overset {\cdot }{P}_{\mathrm{REF}}| \).
However, high values of the surface poloidal field lead to values
of \( \overset {\cdot }{P}_{\mathrm{TOV}} \) which are comparable
to or even greater than \( |\overset {\cdot }{P}_{\mathrm{REF}}| \)
(Figs.~\ref{fig:060_P_timerate}, \ref{fig:090_P_timerate}, and
\ref{fig:132_P_timerate}). For such high values of \( B_{s} \),
\( \overset {\cdot }{P}_{\mathrm{TOV}}/|\overset {\cdot }{P}_{\mathrm{REF}}| \)
depends only weakly on the particular values of \( L_{xx} \) (roughly
speaking, it is independent of \( L_{xx} \)). Consequently, in a
white dwarf which is now in its turn-over phase, the turn-over effects
are negligible if the surface poloidal field is weak; on the other
hand, if the surface poloidal field is strong or even moderate, the
turn-over effects to the time rate of the central period cannot be
neglected.

\acknowledgements{\textbf{ACKNOWLEDGMENTS}}

The research reported here was supported by the Research Committee
of the University of Patras (C. Carath\'eodory's Research Project
1998/1932).

\begin{deluxetable}{ccccccccc}
\tablewidth{0pt}
\tablecaption{Case $M/M_\odot =0.6$:~Results concerning the aligned rotator (part I)\label{tab:060_Aligned_constant_for_Bs}}
\tablehead{
\colhead{$L_{xx}$}& \colhead{$P_{xx}$}& \colhead{$T_{xx}$}& \colhead{$I_{11}$}& \colhead{$I_{33}$}& \colhead{$\left\langle H_{ts}\right\rangle$}& \colhead{$H_{t \left[\mathrm{max}\right]}$}& \colhead{$\left\langle H_t\right\rangle$}& \colhead{$\omega_\mathrm{e}^{-1}$}\\
}
\tabletypesize{\scriptsize}
\startdata
5.150e+48& 141.5427& 8.0177e+46& 1.7858e+50& 1.7336e+50& 4.3306e+09& 2.6760e+11& 1.1623e+11& 3.2106\\
5.500e+48& 132.1198& 9.1358e+46& 1.7868e+50& 1.7353e+50& 4.3314e+09& 2.6739e+11& 1.1613e+11& 3.2105\\
6.000e+48& 120.7721& 1.0856e+47& 1.7884e+50& 1.7380e+50& 4.3326e+09& 2.6706e+11& 1.1600e+11& 3.2105\\
6.500e+48& 111.5137& 1.2719e+47& 1.7900e+50& 1.7409e+50& 4.3339e+09& 2.6670e+11& 1.1585e+11& 3.2104\\
7.000e+48& 103.9030& 1.4725e+47& 1.7918e+50& 1.7440e+50& 4.3353e+09& 2.6632e+11& 1.1569e+11& 3.2091\\
7.500e+48& \phn 97.3565& 1.6870e+47& 1.7938e+50& 1.7474e+50& 4.3367e+09& 2.6591e+11& 1.1552e+11& 3.2091\\
7.900e+48& \phn 92.5301& 1.8687e+47& 1.7954e+50& 1.7503e+50& 4.3380e+09& 2.6556e+11& 1.1537e+11& 3.2090\\
\enddata
\tablecomments{In all tables, parameters are given in cgs units, unless stated otherwise.}
\end{deluxetable}

\begin{deluxetable}{cccccc}
\tablewidth{0pt}
\tablecaption{Case $M/M_\odot =0.6$:~Results concerning the aligned rotator (part II)\label{tab:060_Aligned_variable_for_Bs}}
\tablehead{
\colhead{$B_s$}& \colhead{$\beta _*^p$}& \colhead{$V_{As}$}& \colhead{$t_{As}$}& \colhead{$H_{p \left[\mathrm{max}\right]}$}& \colhead{$\left\langle H_p\right\rangle$}\\
}
\tabletypesize{\scriptsize}
\startdata
\cutinhead{Models with $L_{xx}=$ 5.150e+48}
5.0e+05& 7.7169e+04& 1.1829e+03& 7.3680e+05& 9.8731e+06& 3.4973e+06\\
2.0e+06& 1.9291e+04& 4.7319e+03& 1.8419e+05& 3.9494e+07& 1.3990e+07\\
3.5e+06& 1.1025e+04& 8.2798e+03& 1.0527e+05& 6.9107e+07& 2.4479e+07\\
5.0e+06& 7.7170e+03& 1.1829e+04& 7.3681e+04& 9.8731e+07& 3.4972e+07\\
6.5e+06& 5.9363e+03& 1.5377e+04& 5.6679e+04& 1.2835e+08& 4.5463e+07\\
8.0e+06& 4.8225e+03& 1.8929e+04& 4.6044e+04& 1.5799e+08& 5.5963e+07\\
9.5e+06& 4.0606e+03& 2.2481e+04& 3.8770e+04& 1.8763e+08& 6.6463e+07\\
\cutinhead{Models with $L_{xx}=$ 5.500e+48}
5.0e+05& 7.7143e+04& 1.1829e+03& 7.3695e+05& 9.8679e+06& 3.4956e+06\\
2.0e+06& 1.9286e+04& 4.7317e+03& 1.8424e+05& 3.9471e+07& 1.3982e+07\\
3.5e+06& 1.1021e+04& 8.2798e+03& 1.0529e+05& 6.9070e+07& 2.4467e+07\\
5.0e+06& 7.7151e+03& 1.1828e+04& 7.3703e+04& 9.8668e+07& 3.4952e+07\\
6.5e+06& 5.9345e+03& 1.5377e+04& 5.6692e+04& 1.2827e+08& 4.5440e+07\\
8.0e+06& 4.8206e+03& 1.8930e+04& 4.6051e+04& 1.5791e+08& 5.5940e+07\\
9.5e+06& 4.0606e+03& 2.2473e+04& 3.8791e+04& 1.8747e+08& 6.6409e+07\\
\cutinhead{Models with $L_{xx}=$ 6.000e+48}
5.0e+05& 7.7104e+04& 1.1829e+03& 7.3719e+05& 9.8597e+06& 3.4931e+06\\
2.0e+06& 1.9277e+04& 4.7316e+03& 1.8430e+05& 3.9438e+07& 1.3972e+07\\
3.5e+06& 1.1016e+04& 8.2798e+03& 1.0532e+05& 6.9012e+07& 2.4450e+07\\
5.0e+06& 7.7096e+03& 1.1831e+04& 7.3711e+04& 9.8608e+07& 3.4935e+07\\
6.5e+06& 5.9307e+03& 1.5379e+04& 5.6703e+04& 1.2818e+08& 4.5413e+07\\
8.0e+06& 4.8187e+03& 1.8928e+04& 4.6072e+04& 1.5776e+08& 5.5892e+07\\
9.5e+06& 4.0588e+03& 2.2472e+04& 3.8806e+04& 1.8730e+08& 6.6357e+07\\
\cutinhead{Models with $L_{xx}=$ 6.500e+48}
5.0e+05& 7.7061e+04& 1.1829e+03& 7.3744e+05& 9.8508e+06& 3.4903e+06\\
2.0e+06& 1.9265e+04& 4.7317e+03& 1.8436e+05& 3.9403e+07& 1.3961e+07\\
3.5e+06& 1.1008e+04& 8.2808e+03& 1.0534e+05& 6.8959e+07& 2.4433e+07\\
5.0e+06& 7.7058e+03& 1.1830e+04& 7.3741e+04& 9.8512e+07& 3.4905e+07\\
6.5e+06& 5.9270e+03& 1.5380e+04& 5.6719e+04& 1.2808e+08& 4.5380e+07\\
8.0e+06& 4.8169e+03& 1.8925e+04& 4.6095e+04& 1.5760e+08& 5.5839e+07\\
9.5e+06& 4.0550e+03& 2.2480e+04& 3.8805e+04& 1.8720e+08& 6.6329e+07\\
\cutinhead{Models with $L_{xx}=$ 7.000e+48}
5.0e+05& 7.7015e+04& 1.1829e+03& 7.3771e+05& 9.8415e+06& 3.4874e+06\\
2.0e+06& 1.9254e+04& 4.7316e+03& 1.8443e+05& 3.9365e+07& 1.3949e+07\\
3.5e+06& 1.1003e+04& 8.2800e+03& 1.0539e+05& 6.8886e+07& 2.4411e+07\\
5.0e+06& 7.7021e+03& 1.1828e+04& 7.3777e+04& 9.8407e+07& 3.4871e+07\\
6.5e+06& 5.9233e+03& 1.5380e+04& 5.6738e+04& 1.2796e+08& 4.5343e+07\\
8.0e+06& 4.8131e+03& 1.8928e+04& 4.6104e+04& 1.5747e+08& 5.5802e+07\\
9.5e+06& 4.0532e+03& 2.2477e+04& 3.8825e+04& 1.8700e+08& 6.6264e+07\\
\cutinhead{Models with $L_{xx}=$ 7.500e+48}
5.0e+05& 7.6965e+04& 1.1829e+03& 7.3800e+05& 9.8314e+06& 3.4843e+06\\
2.0e+06& 1.9241e+04& 4.7317e+03& 1.8450e+05& 3.9326e+07& 1.3937e+07\\
3.5e+06& 1.0995e+04& 8.2803e+03& 1.0543e+05& 6.8819e+07& 2.4390e+07\\
5.0e+06& 7.6965e+03& 1.1829e+04& 7.3799e+04& 9.8315e+07& 3.4843e+07\\
6.5e+06& 5.9196e+03& 1.5380e+04& 5.6761e+04& 1.2783e+08& 4.5302e+07\\
8.0e+06& 4.8094e+03& 1.8930e+04& 4.6116e+04& 1.5733e+08& 5.5759e+07\\
9.5e+06& 4.0513e+03& 2.2473e+04& 3.8847e+04& 1.8677e+08& 6.6193e+07\\
\cutinhead{Models with $L_{xx}=$ 7.900e+48}
5.0e+05& 7.6924e+04& 1.1829e+03& 7.3824e+05& 9.8230e+06& 3.4817e+06\\
2.0e+06& 1.9232e+04& 4.7315e+03& 1.8457e+05& 3.9290e+07& 1.3926e+07\\
3.5e+06& 1.0990e+04& 8.2801e+03& 1.0547e+05& 6.8757e+07& 2.4370e+07\\
5.0e+06& 7.6928e+03& 1.1829e+04& 7.3828e+04& 9.8224e+07& 3.4815e+07\\
6.5e+06& 5.9177e+03& 1.5377e+04& 5.6792e+04& 1.2769e+08& 4.5258e+07\\
8.0e+06& 4.8076e+03& 1.8928e+04& 4.6138e+04& 1.5717e+08& 5.5709e+07\\
9.5e+06& 4.0495e+03& 2.2471e+04& 3.8863e+04& 1.8660e+08& 6.6138e+07\\
\enddata
\end{deluxetable}

\begin{deluxetable}{ccccccccc}
\tablewidth{0pt}
\tablecaption{Case $M/M_\odot =0.6$:~Results concerning the perpendicular rotator (part I)\label{tab:060_Perpendicular_constant_for_Bs}}
\tablehead{
\colhead{$L_{xx}$}& \colhead{$P_\mathrm{RR}$}& \colhead{$T_\mathrm{RR}$}& \colhead{$I_{11}$}& \colhead{$I_{33}$}& \colhead{$\left\langle H_{ts}\right\rangle$}& \colhead{$H_{t \left[\mathrm{max}\right]}$}& \colhead{$\left\langle H_t\right\rangle$}& \colhead{$\omega_\mathrm{e}^{-1}$}\\
}
\tabletypesize{\scriptsize}
\startdata
5.150e+48& 218.3990& 7.4081e+46& 1.7901e+50& 1.7210e+50& 4.3247e+09& 2.6924e+11& 1.1691e+11& 1.0000\\
5.500e+48& 204.6816& 8.4418e+46& 1.7917e+50& 1.7210e+50& 4.3247e+09& 2.6924e+11& 1.1691e+11& 1.0000\\
6.000e+48& 187.8799& 1.0033e+47& 1.7941e+50& 1.7210e+50& 4.3247e+09& 2.6924e+11& 1.1691e+11& 1.0000\\
6.500e+48& 173.6831& 1.1757e+47& 1.7968e+50& 1.7210e+50& 4.3247e+09& 2.6924e+11& 1.1691e+11& 1.0000\\
7.000e+48& 161.5330& 1.3614e+47& 1.7996e+50& 1.7210e+50& 4.3247e+09& 2.6924e+11& 1.1691e+11& 1.0000\\
7.500e+48& 151.0202& 1.5602e+47& 1.8027e+50& 1.7210e+50& 4.3247e+09& 2.6924e+11& 1.1691e+11& 1.0000\\
7.900e+48& 143.5799& 1.7286e+47& 1.8053e+50& 1.7210e+50& 4.3247e+09& 2.6924e+11& 1.1691e+11& 1.0000\\
\enddata
\end{deluxetable}

\begin{deluxetable}{cccccc}
\tablewidth{0pt}
\tablecaption{Case $M/M_\odot =0.6$:~Results concerning the perpendicular rotator (part II)\label{tab:060_Perpendicular_variable_for_Bs}}
\tablehead{
\colhead{$B_s$}& \colhead{$\beta _*^p$}& \colhead{$V_{As}$}& \colhead{$t_{As}$}& \colhead{$H_{p \left[\mathrm{max}\right]}$}& \colhead{$\left\langle H_p\right\rangle$}\\
}
\tabletypesize{\scriptsize}
\startdata
\cutinhead{Models with $L_{xx}=$ 5.150e+48}
5.0e+05& 7.6963e+04& 1.1891e+03& 7.3184e+05& 9.9651e+06& 3.5280e+06\\
2.0e+06& 1.9241e+04& 4.7562e+03& 1.8297e+05& 3.9859e+07& 1.4112e+07\\
3.5e+06& 1.0994e+04& 8.3242e+03& 1.0454e+05& 6.9760e+07& 2.4698e+07\\
5.0e+06& 7.6962e+03& 1.1891e+04& 7.3183e+04& 9.9653e+07& 3.5281e+07\\
6.5e+06& 5.9205e+03& 1.5458e+04& 5.6297e+04& 1.2954e+08& 4.5863e+07\\
8.0e+06& 4.8104e+03& 1.9025e+04& 4.5742e+04& 1.5944e+08& 5.6447e+07\\
9.5e+06& 4.0513e+03& 2.2590e+04& 3.8523e+04& 1.8931e+08& 6.7024e+07\\
\cutinhead{Models with $L_{xx}=$ 5.500e+48}
5.0e+05& 7.6910e+04& 1.1899e+03& 7.3134e+05& 9.9720e+06& 3.5305e+06\\
2.0e+06& 1.9228e+04& 4.7595e+03& 1.8284e+05& 3.9887e+07& 1.4122e+07\\
3.5e+06& 1.0986e+04& 8.3300e+03& 1.0447e+05& 6.9809e+07& 2.4715e+07\\
5.0e+06& 7.6905e+03& 1.1900e+04& 7.3129e+04& 9.9727e+07& 3.5307e+07\\
6.5e+06& 5.9166e+03& 1.5468e+04& 5.6261e+04& 1.2963e+08& 4.5893e+07\\
8.0e+06& 4.8066e+03& 1.9040e+04& 4.5706e+04& 1.5956e+08& 5.6492e+07\\
9.6e+06& 4.0474e+03& 2.2611e+04& 3.8487e+04& 1.8949e+08& 6.7087e+07\\
\cutinhead{Models with $L_{xx}=$ 6.000e+48}
5.0e+05& 7.6828e+04& 1.1912e+03& 7.3056e+05& 9.9826e+06& 3.5343e+06\\
2.0e+06& 1.9207e+04& 4.7647e+03& 1.8264e+05& 3.9931e+07& 1.4137e+07\\
3.5e+06& 1.0975e+04& 8.3387e+03& 1.0436e+05& 6.9882e+07& 2.4741e+07\\
5.0e+06& 7.6828e+03& 1.1912e+04& 7.3056e+04& 9.9826e+07& 3.5342e+07\\
6.5e+06& 5.9090e+03& 1.5488e+04& 5.6189e+04& 1.2979e+08& 4.5952e+07\\
8.1e+06& 4.8008e+03& 1.9063e+04& 4.5651e+04& 1.5975e+08& 5.6559e+07\\
9.6e+06& 4.0436e+03& 2.2632e+04& 3.8451e+04& 1.8967e+08& 6.7150e+07\\
\cutinhead{Models with $L_{xx}=$ 6.500e+48}
5.0e+05& 7.6738e+04& 1.1926e+03& 7.2970e+05& 9.9943e+06& 3.5384e+06\\
2.0e+06& 1.9184e+04& 4.7704e+03& 1.8242e+05& 3.9978e+07& 1.4154e+07\\
3.5e+06& 1.0963e+04& 8.3474e+03& 1.0425e+05& 6.9955e+07& 2.4767e+07\\
5.0e+06& 7.6733e+03& 1.1927e+04& 7.2965e+04& 9.9950e+07& 3.5386e+07\\
6.6e+06& 5.9033e+03& 1.5503e+04& 5.6134e+04& 1.2992e+08& 4.5996e+07\\
8.1e+06& 4.7970e+03& 1.9078e+04& 4.5615e+04& 1.5988e+08& 5.6604e+07\\
9.6e+06& 4.0379e+03& 2.2664e+04& 3.8396e+04& 1.8994e+08& 6.7245e+07\\
\cutinhead{Models with $L_{xx}=$ 7.000e+48}
5.0e+05& 7.6642e+04& 1.1941e+03& 7.2879e+05& 1.0007e+07& 3.5428e+06\\
2.0e+06& 1.9161e+04& 4.7761e+03& 1.8220e+05& 4.0026e+07& 1.4171e+07\\
3.5e+06& 1.0948e+04& 8.3590e+03& 1.0411e+05& 7.0052e+07& 2.4801e+07\\
5.0e+06& 7.6638e+03& 1.1941e+04& 7.2875e+04& 1.0007e+08& 3.5430e+07\\
6.6e+06& 5.8957e+03& 1.5523e+04& 5.6062e+04& 1.3009e+08& 4.6056e+07\\
8.1e+06& 4.7894e+03& 1.9108e+04& 4.5542e+04& 1.6013e+08& 5.6694e+07\\
9.6e+06& 4.0341e+03& 2.2686e+04& 3.8360e+04& 1.9012e+08& 6.7309e+07\\
\cutinhead{Models with $L_{xx}=$ 7.500e+48}
5.1e+05& 7.6540e+04& 1.1957e+03& 7.2782e+05& 1.0020e+07& 3.5476e+06\\
2.0e+06& 1.9135e+04& 4.7828e+03& 1.8195e+05& 4.0082e+07& 1.4191e+07\\
3.5e+06& 1.0935e+04& 8.3692e+03& 1.0398e+05& 7.0138e+07& 2.4832e+07\\
5.1e+06& 7.6542e+03& 1.1956e+04& 7.2784e+04& 1.0020e+08& 3.5475e+07\\
6.6e+06& 5.8880e+03& 1.5543e+04& 5.5989e+04& 1.3026e+08& 4.6116e+07\\
8.1e+06& 4.7837e+03& 1.9131e+04& 4.5488e+04& 1.6033e+08& 5.6762e+07\\
9.6e+06& 4.0284e+03& 2.2718e+04& 3.8306e+04& 1.9039e+08& 6.7405e+07\\
\cutinhead{Models with $L_{xx}=$ 7.900e+48}
5.1e+05& 7.6452e+04& 1.1970e+03& 7.2698e+05& 1.0032e+07& 3.5516e+06\\
2.0e+06& 1.9114e+04& 4.7880e+03& 1.8175e+05& 4.0126e+07& 1.4206e+07\\
3.5e+06& 1.0922e+04& 8.3794e+03& 1.0385e+05& 7.0223e+07& 2.4862e+07\\
5.1e+06& 7.6447e+03& 1.1971e+04& 7.2693e+04& 1.0032e+08& 3.5519e+07\\
6.6e+06& 5.8804e+03& 1.5563e+04& 5.5917e+04& 1.3042e+08& 4.6175e+07\\
8.1e+06& 4.7779e+03& 1.9154e+04& 4.5433e+04& 1.6052e+08& 5.6830e+07\\
9.6e+06& 4.0245e+03& 2.2740e+04& 3.8269e+04& 1.9057e+08& 6.7468e+07\\
\enddata
\end{deluxetable}

\begin{deluxetable}{ccccccccccc}
\tablewidth{0pt}
\tablecaption{Case $M/M_\odot =0.6$:~Results concerning the turn-over phase\label{tab:060_Turn_Over}}
\tablehead{
\colhead{$B_s$}& \colhead{$f$}& \colhead{$t_\mathrm{TOV}$}& \colhead{$\overset{.}{P}_\mathrm{TOV}$}& \colhead{$N_{xx}$}& \colhead{$t_\mathrm{now}$}& \colhead{$\overset{.}{T}$}\\
}
\tabletypesize{\scriptsize}
\startdata
\cutinhead{$L_{xx}=$ 5.150e+48, $A_r=$ 1.0065, $\chi _\mathrm{now}=$ 6.5298, $F_{r\left[\mathrm{now}\right]}=$ 0.9938}
5.0e+05& 3.9951e+23& 7.3392e+09& 3.3207e-16& 1.6352e+15& 4.3666e+07& 2.6338e+28\\
2.0e+06& 1.5980e+24& 4.4475e+08& 5.4797e-15& 9.9090e+13& 2.6461e+06& 4.3463e+29\\
3.5e+06& 2.7966e+24& 1.4995e+08& 1.6252e-14& 3.3410e+13& 8.9218e+05& 1.2891e+30\\
5.0e+06& 3.9951e+24& 7.0972e+07& 3.4339e-14& 1.5813e+13& 4.2227e+05& 2.7236e+30\\
6.5e+06& 5.1936e+24& 4.1868e+07& 5.8208e-14& 9.3284e+12& 2.4911e+05& 4.6169e+30\\
8.0e+06& 6.3921e+24& 2.7635e+07& 8.8189e-14& 6.1571e+12& 1.6442e+05& 6.9948e+30\\
9.5e+06& 7.5907e+24& 1.9571e+07& 1.2452e-13& 4.3606e+12& 1.1645e+05& 9.8767e+30\\
\cutinhead{$L_{xx}=$ 5.500e+48, $A_r=$ 1.0074, $\chi _\mathrm{now}=$ 32.2461, $F_{r\left[\mathrm{now}\right]}=$ 0.8460}
5.0e+05& 3.9979e+23& 5.5981e+09& 4.1102e-16& 1.3362e+15& 7.6225e+08& 3.9311e+28\\
2.0e+06& 1.5992e+24& 3.3927e+08& 6.7819e-15& 8.0982e+13& 4.6196e+07& 6.4864e+29\\
3.5e+06& 2.7985e+24& 1.1438e+08& 2.0117e-14& 2.7301e+13& 1.5574e+07& 1.9240e+30\\
5.0e+06& 3.9979e+24& 5.4146e+07& 4.2495e-14& 1.2924e+13& 7.3726e+06& 4.0643e+30\\
6.5e+06& 5.1972e+24& 3.1937e+07& 7.2045e-14& 7.6232e+12& 4.3487e+06& 6.8906e+30\\
8.0e+06& 6.3966e+24& 2.1077e+07& 1.0917e-13& 5.0310e+12& 2.8699e+06& 1.0441e+31\\
9.5e+06& 7.5960e+24& 1.4939e+07& 1.5403e-13& 3.5657e+12& 2.0341e+06& 1.4731e+31\\
\cutinhead{$L_{xx}=$ 6.000e+48, $A_r=$ 1.0087, $\chi _\mathrm{now}=$ 51.9417, $F_{r\left[\mathrm{now}\right]}=$ 0.6166}
5.0e+05& 4.0022e+23& 3.9250e+09& 5.4216e-16& 1.0249e+15& 1.2416e+09& 6.6504e+28\\
2.0e+06& 1.6009e+24& 2.3789e+08& 8.9453e-15& 6.2117e+13& 7.5250e+07& 1.0973e+30\\
3.5e+06& 2.8015e+24& 8.0195e+07& 2.6535e-14& 2.0941e+13& 2.5368e+07& 3.2549e+30\\
5.0e+06& 4.0022e+24& 3.7948e+07& 5.6076e-14& 9.9089e+12& 1.2004e+07& 6.8786e+30\\
6.5e+06& 5.2028e+24& 2.2387e+07& 9.5054e-14& 5.8457e+12& 7.0816e+06& 1.1660e+31\\
8.0e+06& 6.4035e+24& 1.4782e+07& 1.4396e-13& 3.8598e+12& 4.6758e+06& 1.7659e+31\\
9.5e+06& 7.6041e+24& 1.0475e+07& 2.0315e-13& 2.7352e+12& 3.3135e+06& 2.4919e+31\\
\cutinhead{$L_{xx}=$ 6.500e+48, $A_r=$ 1.0102, $\chi _\mathrm{now}=$ 66.4632, $F_{r\left[\mathrm{now}\right]}=$ 0.3995}
5.0e+05& 4.0068e+23& 2.8474e+09& 6.9234e-16& 8.0525e+14& 1.3963e+09& 1.0715e+29\\
2.0e+06& 1.6027e+24& 1.7257e+08& 1.1424e-14& 4.8803e+13& 8.4625e+07& 1.7679e+30\\
3.5e+06& 2.8048e+24& 5.8165e+07& 3.3893e-14& 1.6449e+13& 2.8522e+07& 5.2453e+30\\
5.0e+06& 4.0068e+24& 2.7534e+07& 7.1598e-14& 7.7866e+12& 1.3502e+07& 1.1080e+31\\
6.5e+06& 5.2089e+24& 1.6239e+07& 1.2140e-13& 4.5923e+12& 7.9630e+06& 1.8788e+31\\
8.0e+06& 6.4109e+24& 1.0727e+07& 1.8377e-13& 3.0337e+12& 5.2604e+06& 2.8441e+31\\
9.5e+06& 7.6130e+24& 7.5938e+06& 2.5960e-13& 2.1475e+12& 3.7238e+06& 4.0176e+31\\
\cutinhead{$L_{xx}=$ 7.000e+48, $A_r=$ 1.0118, $\chi _\mathrm{now}=$ 77.2625, $F_{r\left[\mathrm{now}\right]}=$ 0.2206}
5.0e+05& 4.0118e+23& 2.1235e+09& 8.6058e-16& 6.4451e+14& 1.4038e+09& 1.6584e+29\\
2.0e+06& 1.6047e+24& 1.2871e+08& 1.4199e-14& 3.9064e+13& 8.5082e+07& 2.7361e+30\\
3.5e+06& 2.8083e+24& 4.3386e+07& 4.2121e-14& 1.3168e+13& 2.8681e+07& 8.1168e+30\\
5.0e+06& 4.0118e+24& 2.0539e+07& 8.8974e-14& 6.2338e+12& 1.3577e+07& 1.7146e+31\\
6.5e+06& 5.2154e+24& 1.2110e+07& 1.5091e-13& 3.6754e+12& 8.0052e+06& 2.9081e+31\\
8.0e+06& 6.4190e+24& 7.9974e+06& 2.2850e-13& 2.4273e+12& 5.2868e+06& 4.4033e+31\\
9.5e+06& 7.6225e+24& 5.6648e+06& 3.2259e-13& 1.7194e+12& 3.7448e+06& 6.2165e+31\\
\cutinhead{$L_{xx}=$ 7.500e+48, $A_r=$ 1.0136, $\chi _\mathrm{now}=$ 84.9531, $F_{r\left[\mathrm{now}\right]}=$ 0.0880}
5.0e+05& 4.0172e+23& 1.6251e+09& 1.0471e-15& 5.2640e+14& 1.3519e+09& 2.4750e+29\\
2.0e+06& 1.6069e+24& 9.8489e+07& 1.7278e-14& 3.1903e+13& 8.1934e+07& 4.0838e+30\\
3.5e+06& 2.8121e+24& 3.3200e+07& 5.1256e-14& 1.0754e+13& 2.7619e+07& 1.2115e+31\\
5.0e+06& 4.0172e+24& 1.5716e+07& 1.0828e-13& 5.0906e+12& 1.3074e+07& 2.5593e+31\\
6.5e+06& 5.2224e+24& 9.2674e+06& 1.8362e-13& 3.0019e+12& 7.7096e+06& 4.3401e+31\\
8.0e+06& 6.4276e+24& 6.1187e+06& 2.7811e-13& 1.9820e+12& 5.0902e+06& 6.5734e+31\\
9.5e+06& 7.6327e+24& 4.3368e+06& 3.9238e-13& 1.4048e+12& 3.6078e+06& 9.2744e+31\\
\cutinhead{$L_{xx}=$ 7.900e+48, $A_r=$ 1.0150, $\chi _\mathrm{now}=$ 89.2287, $F_{r\left[\mathrm{now}\right]}=$ 0.0135}
5.0e+05& 4.0218e+23& 1.3288e+09& 1.2182e-15& 4.5290e+14& 1.2877e+09& 3.3443e+29\\
2.0e+06& 1.6087e+24& 8.0547e+07& 2.0097e-14& 2.7452e+13& 7.8054e+07& 5.5173e+30\\
3.5e+06& 2.8152e+24& 2.7150e+07& 5.9624e-14& 9.2532e+12& 2.6310e+07& 1.6368e+31\\
5.0e+06& 4.0218e+24& 1.2853e+07& 1.2595e-13& 4.3804e+12& 1.2455e+07& 3.4577e+31\\
6.5e+06& 5.2283e+24& 7.5816e+06& 2.1351e-13& 2.5840e+12& 7.3470e+06& 5.8616e+31\\
8.0e+06& 6.4348e+24& 5.0050e+06& 3.2343e-13& 1.7058e+12& 4.8501e+06& 8.8791e+31\\
9.5e+06& 7.6414e+24& 3.5469e+06& 4.5639e-13& 1.2088e+12& 3.4371e+06& 1.2529e+32\\
\enddata
\tablecomments{$t_\mathrm{TOV}$ and $t_\mathrm{now}$ are given in years. $\chi_\mathrm{now}$ is given in arcdegrees.}
\end{deluxetable}

\begin{deluxetable}{ccccccccc}
\tablewidth{0pt}
\tablecaption{Case $M/M_\odot =0.9$:~Results concerning the aligned rotator (part I)\label{tab:090_Aligned_constant_for_Bs}}
\tablehead{
\colhead{$L_{xx}$}& \colhead{$P_{xx}$}& \colhead{$T_{xx}$}& \colhead{$I_{11}$}& \colhead{$I_{33}$}& \colhead{$\left\langle H_{ts}\right\rangle$}& \colhead{$H_{t \left[\mathrm{max}\right]}$}& \colhead{$\left\langle H_t\right\rangle$}& \colhead{$\omega_\mathrm{e}^{-1}$}\\
}
\tabletypesize{\scriptsize}
\startdata
1.620e+49& 32.9038& 1.0425e+48& 1.3595e+50& 1.3358e+50& 2.7897e+09& 7.9651e+11& 3.2773e+11& 3.6838\\
1.650e+49& 31.9878& 1.0799e+48& 1.3608e+50& 1.3378e+50& 2.7908e+09& 7.9549e+11& 3.2733e+11& 3.6830\\
1.850e+49& 28.6012& 1.3434e+48& 1.3705e+50& 1.3519e+50& 2.7979e+09& 7.8823e+11& 3.2454e+11& 3.6777\\
2.050e+49& 26.0253& 1.6305e+48& 1.3812e+50& 1.3676e+50& 2.8057e+09& 7.8033e+11& 3.2149e+11& 3.6724\\
2.250e+49& 24.0878& 1.9393e+48& 1.3930e+50& 1.3848e+50& 2.8142e+09& 7.7184e+11& 3.1821e+11& 3.6664\\
2.450e+49& 22.6166& 2.2683e+48& 1.4057e+50& 1.4035e+50& 2.8234e+09& 7.6283e+11& 3.1473e+11& 3.6599\\
2.515e+49& 22.1804& 2.3792e+48& 1.4100e+50& 1.4099e+50& 2.8265e+09& 7.5981e+11& 3.1356e+11& 3.6575\\
\enddata
\end{deluxetable}

\begin{deluxetable}{cccccc}
\tablewidth{0pt}
\tablecaption{Case $M/M_\odot =0.9$:~Results concerning the aligned rotator (part II)\label{tab:090_Aligned_variable_for_Bs}}
\tablehead{
\colhead{$B_s$}& \colhead{$\beta _*^p$}& \colhead{$V_{As}$}& \colhead{$t_{As}$}& \colhead{$H_{p \left[\mathrm{max}\right]}$}& \colhead{$\left\langle H_p\right\rangle$}\\
}
\tabletypesize{\scriptsize}
\startdata
\cutinhead{Models with $L_{xx}=$ 1.620e+49}
5.0e+05& 1.0210e+05& 1.2872e+03& 5.0307e+05& 2.3724e+07& 7.5225e+06\\
2.0e+06& 2.5526e+04& 5.1488e+03& 1.2577e+05& 9.4895e+07& 3.0089e+07\\
3.5e+06& 1.4586e+04& 9.0102e+03& 7.1870e+04& 1.6606e+08& 5.2655e+07\\
5.0e+06& 1.0209e+04& 1.2873e+04& 5.0303e+04& 2.3726e+08& 7.5231e+07\\
6.5e+06& 7.8548e+03& 1.6732e+04& 3.8702e+04& 3.0838e+08& 9.7781e+07\\
8.0e+06& 6.3815e+03& 2.0595e+04& 3.1443e+04& 3.7958e+08& 1.2036e+08\\
9.5e+06& 5.3738e+03& 2.4457e+04& 2.6478e+04& 4.5075e+08& 1.4293e+08\\
\cutinhead{Models with $L_{xx}=$ 1.650e+49}
5.0e+05& 1.0207e+05& 1.2872e+03& 5.0326e+05& 2.3699e+07& 7.5155e+06\\
2.0e+06& 2.5518e+04& 5.1489e+03& 1.2581e+05& 9.4796e+07& 3.0062e+07\\
3.5e+06& 1.4583e+04& 9.0101e+03& 7.1897e+04& 1.6588e+08& 5.2606e+07\\
5.0e+06& 1.0207e+04& 1.2872e+04& 5.0325e+04& 2.3699e+08& 7.5156e+07\\
6.5e+06& 7.8511e+03& 1.6735e+04& 3.8708e+04& 3.0811e+08& 9.7711e+07\\
8.0e+06& 6.3796e+03& 2.0596e+04& 3.1454e+04& 3.7918e+08& 1.2025e+08\\
9.5e+06& 5.3719e+03& 2.4459e+04& 2.6485e+04& 4.5031e+08& 1.4281e+08\\
\cutinhead{Models with $L_{xx}=$ 1.850e+49}
5.0e+05& 1.0188e+05& 1.2872e+03& 5.0460e+05& 2.3519e+07& 7.4665e+06\\
2.0e+06& 2.5470e+04& 5.1489e+03& 1.2615e+05& 9.4074e+07& 2.9866e+07\\
3.5e+06& 1.4555e+04& 9.0102e+03& 7.2088e+04& 1.6462e+08& 5.2264e+07\\
5.0e+06& 1.0189e+04& 1.2871e+04& 5.0464e+04& 2.3517e+08& 7.4659e+07\\
6.5e+06& 7.8362e+03& 1.6735e+04& 3.8812e+04& 3.0577e+08& 9.7073e+07\\
8.0e+06& 6.3666e+03& 2.0598e+04& 3.1533e+04& 3.7635e+08& 1.1948e+08\\
9.5e+06& 5.3626e+03& 2.4455e+04& 2.6561e+04& 4.4681e+08& 1.4185e+08\\
\cutinhead{Models with $L_{xx}=$ 2.050e+49}
5.0e+05& 1.0167e+05& 1.2872e+03& 5.0608e+05& 2.3322e+07& 7.4130e+06\\
2.0e+06& 2.5416e+04& 5.1490e+03& 1.2652e+05& 9.3290e+07& 2.9653e+07\\
3.5e+06& 1.4523e+04& 9.0108e+03& 7.2294e+04& 1.6326e+08& 5.1893e+07\\
5.0e+06& 1.0166e+04& 1.2872e+04& 5.0607e+04& 2.3322e+08& 7.4131e+07\\
6.5e+06& 7.8213e+03& 1.6732e+04& 3.8934e+04& 3.0315e+08& 9.6358e+07\\
8.0e+06& 6.3536e+03& 2.0597e+04& 3.1627e+04& 3.7318e+08& 1.1862e+08\\
9.5e+06& 5.3514e+03& 2.4454e+04& 2.6639e+04& 4.4306e+08& 1.4083e+08\\
\cutinhead{Models with $L_{xx}=$ 2.250e+49}
5.0e+05& 1.0143e+05& 1.2872e+03& 5.0769e+05& 2.3111e+07& 7.3556e+06\\
2.0e+06& 2.5358e+04& 5.1487e+03& 1.2692e+05& 9.2443e+07& 2.9422e+07\\
3.5e+06& 1.4490e+04& 9.0108e+03& 7.2524e+04& 1.6178e+08& 5.1491e+07\\
5.0e+06& 1.0142e+04& 1.2873e+04& 5.0764e+04& 2.3113e+08& 7.3562e+07\\
6.5e+06& 7.8027e+03& 1.6733e+04& 3.9055e+04& 3.0043e+08& 9.5618e+07\\
8.0e+06& 6.3387e+03& 2.0598e+04& 3.1727e+04& 3.6982e+08& 1.1770e+08\\
9.5e+06& 5.3384e+03& 2.4457e+04& 2.6720e+04& 4.3912e+08& 1.3976e+08\\
\cutinhead{Models with $L_{xx}=$ 2.450e+49}
5.0e+05& 1.0118e+05& 1.2872e+03& 5.0941e+05& 2.2888e+07& 7.2945e+06\\
2.0e+06& 2.5295e+04& 5.1487e+03& 1.2736e+05& 9.1548e+07& 2.9177e+07\\
3.5e+06& 1.4454e+04& 9.0103e+03& 7.2775e+04& 1.6021e+08& 5.1060e+07\\
5.0e+06& 1.0118e+04& 1.2872e+04& 5.0943e+04& 2.2887e+08& 7.2943e+07\\
6.5e+06& 7.7822e+03& 1.6735e+04& 3.9182e+04& 2.9756e+08& 9.4836e+07\\
8.0e+06& 6.3237e+03& 2.0595e+04& 3.1839e+04& 3.6619e+08& 1.1671e+08\\
9.5e+06& 5.3254e+03& 2.4456e+04& 2.6813e+04& 4.3484e+08& 1.3859e+08\\
\cutinhead{Models with $L_{xx}=$ 2.515e+49}
5.0e+05& 1.0109e+05& 1.2872e+03& 5.0999e+05& 2.2813e+07& 7.2740e+06\\
2.0e+06& 2.5272e+04& 5.1489e+03& 1.2750e+05& 9.1251e+07& 2.9096e+07\\
3.5e+06& 1.4441e+04& 9.0107e+03& 7.2855e+04& 1.5969e+08& 5.0919e+07\\
5.0e+06& 1.0109e+04& 1.2873e+04& 5.0998e+04& 2.2813e+08& 7.2742e+07\\
6.5e+06& 7.7766e+03& 1.6733e+04& 3.9233e+04& 2.9655e+08& 9.4556e+07\\
8.0e+06& 6.3182e+03& 2.0595e+04& 3.1875e+04& 3.6500e+08& 1.1638e+08\\
9.5e+06& 5.3198e+03& 2.4461e+04& 2.6838e+04& 4.3350e+08& 1.3823e+08\\
\enddata
\end{deluxetable}

\begin{deluxetable}{ccccccccc}
\tablewidth{0pt}
\tablecaption{Case $M/M_\odot =0.9$:~Results concerning the perpendicular rotator (part I)\label{tab:090_Perpendicular_constant_for_Bs}}
\tablehead{
\colhead{$L_{xx}$}& \colhead{$P_\mathrm{RR}$}& \colhead{$T_\mathrm{RR}$}& \colhead{$I_{11}$}& \colhead{$I_{33}$}& \colhead{$\left\langle H_{ts}\right\rangle$}& \colhead{$H_{t \left[\mathrm{max}\right]}$}& \colhead{$\left\langle H_t\right\rangle$}& \colhead{$\omega_\mathrm{e}^{-1}$}\\
}
\tabletypesize{\scriptsize}
\startdata
1.620e+49& 53.1206& 9.5808e+47& 1.3696e+50& 1.2828e+50& 2.7624e+09& 8.2518e+11& 3.3875e+11& 1.0000\\
1.650e+49& 52.2320& 9.9242e+47& 1.3716e+50& 1.2828e+50& 2.7624e+09& 8.2518e+11& 3.3875e+11& 1.0000\\
1.850e+49& 46.9921& 1.2368e+48& 1.3836e+50& 1.2828e+50& 2.7624e+09& 8.2518e+11& 3.3875e+11& 1.0000\\
2.050e+49& 42.8154& 1.5042e+48& 1.3969e+50& 1.2828e+50& 2.7624e+09& 8.2518e+11& 3.3875e+11& 1.0000\\
2.250e+49& 39.4104& 1.7936e+48& 1.4113e+50& 1.2828e+50& 2.7624e+09& 8.2518e+11& 3.3875e+11& 1.0000\\
2.450e+49& 36.5795& 2.1042e+48& 1.4263e+50& 1.2828e+50& 2.7624e+09& 8.2518e+11& 3.3875e+11& 1.0000\\
2.515e+49& 35.7638& 2.2092e+48& 1.4315e+50& 1.2828e+50& 2.7624e+09& 8.2518e+11& 3.3875e+11& 1.0000\\
\enddata
\end{deluxetable}

\begin{deluxetable}{cccccc}
\tablewidth{0pt}
\tablecaption{Case $M/M_\odot =0.9$:~Results concerning the perpendicular rotator (part II)\label{tab:090_Perpendicular_variable_for_Bs}}
\tablehead{
\colhead{$B_s$}& \colhead{$\beta _*^p$}& \colhead{$V_{As}$}& \colhead{$t_{As}$}& \colhead{$H_{p \left[\mathrm{max}\right]}$}& \colhead{$\left\langle H_p\right\rangle$}\\
}
\tabletypesize{\scriptsize}
\startdata
\cutinhead{Models with $L_{xx}=$ 1.620e+49}
5.1e+05& 9.9861e+04& 1.3256e+03& 4.8349e+05& 2.5165e+07& 7.9456e+06\\
2.1e+06& 2.4965e+04& 5.3025e+03& 1.2087e+05& 1.0066e+08& 3.1783e+07\\
3.6e+06& 1.4265e+04& 9.2799e+03& 6.9067e+04& 1.7617e+08& 5.5623e+07\\
5.1e+06& 9.9869e+03& 1.3255e+04& 4.8353e+04& 2.5163e+08& 7.9450e+07\\
6.7e+06& 7.6809e+03& 1.7235e+04& 3.7188e+04& 3.2718e+08& 1.0330e+08\\
8.2e+06& 6.2409e+03& 2.1211e+04& 3.0216e+04& 4.0267e+08& 1.2714e+08\\
9.8e+06& 5.2567e+03& 2.5183e+04& 2.5451e+04& 4.7806e+08& 1.5094e+08\\
\cutinhead{Models with $L_{xx}=$ 1.650e+49}
5.2e+05& 9.9754e+04& 1.3270e+03& 4.8297e+05& 2.5192e+07& 7.9542e+06\\
2.1e+06& 2.4939e+04& 5.3082e+03& 1.2074e+05& 1.0077e+08& 3.1817e+07\\
3.6e+06& 1.4250e+04& 9.2898e+03& 6.8993e+04& 1.7635e+08& 5.5682e+07\\
5.2e+06& 9.9755e+03& 1.3270e+04& 4.8298e+04& 2.5192e+08& 7.9541e+07\\
6.7e+06& 7.6733e+03& 1.7252e+04& 3.7151e+04& 3.2750e+08& 1.0341e+08\\
8.2e+06& 6.2352e+03& 2.1231e+04& 3.0188e+04& 4.0304e+08& 1.2726e+08\\
9.8e+06& 5.2510e+03& 2.5210e+04& 2.5423e+04& 4.7858e+08& 1.5111e+08\\
\cutinhead{Models with $L_{xx}=$ 1.850e+49}
5.2e+05& 9.8996e+04& 1.3372e+03& 4.7930e+05& 2.5385e+07& 8.0151e+06\\
2.1e+06& 2.4750e+04& 5.3487e+03& 1.1983e+05& 1.0154e+08& 3.2059e+07\\
3.6e+06& 1.4143e+04& 9.3600e+03& 6.8476e+04& 1.7769e+08& 5.6103e+07\\
5.2e+06& 9.8992e+03& 1.3373e+04& 4.7928e+04& 2.5386e+08& 8.0154e+07\\
6.8e+06& 7.6142e+03& 1.7386e+04& 3.6865e+04& 3.3004e+08& 1.0421e+08\\
8.3e+06& 6.1875e+03& 2.1395e+04& 2.9958e+04& 4.0615e+08& 1.2824e+08\\
9.9e+06& 5.2109e+03& 2.5404e+04& 2.5229e+04& 4.8226e+08& 1.5227e+08\\
\cutinhead{Models with $L_{xx}=$ 2.050e+49}
5.2e+05& 9.8169e+04& 1.3485e+03& 4.7530e+05& 2.5599e+07& 8.0826e+06\\
2.1e+06& 2.4542e+04& 5.3940e+03& 1.1882e+05& 1.0240e+08& 3.2331e+07\\
3.7e+06& 1.4025e+04& 9.4389e+03& 6.7903e+04& 1.7918e+08& 5.6576e+07\\
5.2e+06& 9.8172e+03& 1.3484e+04& 4.7531e+04& 2.5598e+08& 8.0824e+07\\
6.8e+06& 7.5512e+03& 1.7531e+04& 3.6560e+04& 3.3280e+08& 1.0508e+08\\
8.4e+06& 6.1360e+03& 2.1574e+04& 2.9708e+04& 4.0955e+08& 1.2931e+08\\
1.0e+07& 5.1670e+03& 2.5620e+04& 2.5017e+04& 4.8635e+08& 1.5356e+08\\
\cutinhead{Models with $L_{xx}=$ 2.250e+49}
5.3e+05& 9.7280e+04& 1.3608e+03& 4.7100e+05& 2.5833e+07& 8.1564e+06\\
2.1e+06& 2.4321e+04& 5.4430e+03& 1.1775e+05& 1.0333e+08& 3.2625e+07\\
3.7e+06& 1.3897e+04& 9.5257e+03& 6.7284e+04& 1.8083e+08& 5.7096e+07\\
5.3e+06& 9.7275e+03& 1.3609e+04& 4.7097e+04& 2.5834e+08& 8.1569e+07\\
6.9e+06& 7.4826e+03& 1.7692e+04& 3.6228e+04& 3.3585e+08& 1.0604e+08\\
8.5e+06& 6.0807e+03& 2.1770e+04& 2.9440e+04& 4.1328e+08& 1.3049e+08\\
1.0e+07& 5.1194e+03& 2.5858e+04& 2.4786e+04& 4.9088e+08& 1.5499e+08\\
\cutinhead{Models with $L_{xx}=$ 2.450e+49}
5.3e+05& 9.6336e+04& 1.3741e+03& 4.6643e+05& 2.6086e+07& 8.2364e+06\\
2.1e+06& 2.4084e+04& 5.4965e+03& 1.1661e+05& 1.0434e+08& 3.2945e+07\\
3.7e+06& 1.3762e+04& 9.6194e+03& 6.6629e+04& 1.8261e+08& 5.7658e+07\\
5.3e+06& 9.6341e+03& 1.3741e+04& 4.6645e+04& 2.6085e+08& 8.2360e+07\\
6.9e+06& 7.4101e+03& 1.7865e+04& 3.5877e+04& 3.3913e+08& 1.0708e+08\\
8.5e+06& 6.0215e+03& 2.1984e+04& 2.9154e+04& 4.1734e+08& 1.3177e+08\\
1.0e+07& 5.0698e+03& 2.6111e+04& 2.4546e+04& 4.9569e+08& 1.5651e+08\\
\cutinhead{Models with $L_{xx}=$ 2.515e+49}
5.4e+05& 9.6019e+04& 1.3787e+03& 4.6489e+05& 2.6172e+07& 8.2636e+06\\
2.1e+06& 2.4004e+04& 5.5148e+03& 1.1622e+05& 1.0469e+08& 3.3055e+07\\
3.7e+06& 1.3718e+04& 9.6502e+03& 6.6416e+04& 1.8320e+08& 5.7842e+07\\
5.4e+06& 9.6016e+03& 1.3787e+04& 4.6488e+04& 2.6173e+08& 8.2638e+07\\
7.0e+06& 7.3853e+03& 1.7925e+04& 3.5757e+04& 3.4027e+08& 1.0744e+08\\
8.6e+06& 6.0006e+03& 2.2061e+04& 2.9053e+04& 4.1880e+08& 1.3223e+08\\
1.0e+07& 5.0545e+03& 2.6190e+04& 2.4472e+04& 4.9718e+08& 1.5698e+08\\
\enddata
\end{deluxetable}

\begin{deluxetable}{ccccccccccc}
\tablewidth{0pt}
\tablecaption{Case $M/M_\odot =0.9$:~Results concerning the turn-over phase\label{tab:090_Turn_Over}}
\tablehead{
\colhead{$B_s$}& \colhead{$f$}& \colhead{$t_\mathrm{TOV}$}& \colhead{$\overset{.}{P}_\mathrm{TOV}$}& \colhead{$N_{xx}$}& \colhead{$t_\mathrm{now}$}& \colhead{$\overset{.}{T}$}\\
}
\tabletypesize{\scriptsize}
\startdata
\cutinhead{$L_{xx}=$ 1.620e+49, $A_r=$ 1.0352, $\chi _\mathrm{now}=$ 5.7116, $F_{r\left[\mathrm{now}\right]}=$ 0.9953}
5.0e+05& 2.2252e+23& 8.5891e+08& 7.4638e-16& 8.2321e+14& 4.0878e+06& 3.1183e+30\\
2.0e+06& 8.9010e+23& 5.5703e+07& 1.1509e-14& 5.3388e+13& 2.6510e+05& 4.8082e+31\\
3.5e+06& 1.5577e+24& 1.8184e+07& 3.5254e-14& 1.7428e+13& 8.6542e+04& 1.4729e+32\\
5.0e+06& 2.2252e+24& 8.9203e+06& 7.1866e-14& 8.5495e+12& 4.2454e+04& 3.0025e+32\\
6.5e+06& 2.8928e+24& 5.0956e+06& 1.2581e-13& 4.8838e+12& 2.4251e+04& 5.2562e+32\\
8.0e+06& 3.5604e+24& 3.4816e+06& 1.8413e-13& 3.3369e+12& 1.6570e+04& 7.6928e+32\\
9.5e+06& 4.2280e+24& 2.3813e+06& 2.6921e-13& 2.2823e+12& 1.1333e+04& 1.1247e+33\\
\cutinhead{$L_{xx}=$ 1.650e+49, $A_r=$ 1.0368, $\chi _\mathrm{now}=$ 18.6915, $F_{r\left[\mathrm{now}\right]}=$ 0.9475}
5.0e+05& 2.2276e+23& 7.8942e+08& 8.1317e-16& 7.7827e+14& 3.9469e+07& 3.5158e+30\\
2.0e+06& 8.9105e+23& 5.1193e+07& 1.2540e-14& 5.0470e+13& 2.5595e+06& 5.4215e+31\\
3.5e+06& 1.5593e+24& 1.6713e+07& 3.8410e-14& 1.6477e+13& 8.3561e+05& 1.6606e+32\\
5.0e+06& 2.2276e+24& 8.1998e+06& 7.8287e-14& 8.0840e+12& 4.0997e+05& 3.3847e+32\\
6.5e+06& 2.8959e+24& 4.6813e+06& 1.3713e-13& 4.6152e+12& 2.3405e+05& 5.9287e+32\\
8.0e+06& 3.5642e+24& 3.1997e+06& 2.0062e-13& 3.1545e+12& 1.5998e+05& 8.6739e+32\\
9.5e+06& 4.2325e+24& 2.1882e+06& 2.9336e-13& 2.1573e+12& 1.0941e+05& 1.2683e+33\\
\cutinhead{$L_{xx}=$ 1.850e+49, $A_r=$ 1.0458, $\chi _\mathrm{now}=$ 46.6186, $F_{r\left[\mathrm{now}\right]}=$ 0.6871}
5.0e+05& 2.2447e+23& 5.1969e+08& 1.1222e-15& 5.7301e+14& 1.2430e+08& 6.5064e+30\\
2.0e+06& 8.9787e+23& 3.3702e+07& 1.7304e-14& 3.7160e+13& 8.0610e+06& 1.0033e+32\\
3.5e+06& 1.5713e+24& 1.1002e+07& 5.3004e-14& 1.2131e+13& 2.6316e+06& 3.0732e+32\\
5.0e+06& 2.2447e+24& 5.3990e+06& 1.0801e-13& 5.9530e+12& 1.2914e+06& 6.2628e+32\\
6.5e+06& 2.9181e+24& 3.0819e+06& 1.8922e-13& 3.3982e+12& 7.3715e+05& 1.0971e+33\\
8.0e+06& 3.5915e+24& 2.1059e+06& 2.7692e-13& 2.3220e+12& 5.0370e+05& 1.6056e+33\\
9.5e+06& 4.2649e+24& 1.4411e+06& 4.0467e-13& 1.5890e+12& 3.4468e+05& 2.3463e+33\\
\cutinhead{$L_{xx}=$ 2.050e+49, $A_r=$ 1.0559, $\chi _\mathrm{now}=$ 65.6329, $F_{r\left[\mathrm{now}\right]}=$ 0.4127}
5.0e+05& 2.2636e+23& 3.5785e+08& 1.4878e-15& 4.3363e+14& 1.4865e+08& 1.1190e+31\\
2.0e+06& 9.0543e+23& 2.3206e+07& 2.2943e-14& 2.8120e+13& 9.6400e+06& 1.7255e+32\\
3.5e+06& 1.5845e+24& 7.5752e+06& 7.0284e-14& 9.1792e+12& 3.1468e+06& 5.2861e+32\\
5.0e+06& 2.2636e+24& 3.7170e+06& 1.4324e-13& 4.5041e+12& 1.5441e+06& 1.0773e+33\\
6.5e+06& 2.9426e+24& 2.1231e+06& 2.5077e-13& 2.5727e+12& 8.8195e+05& 1.8861e+33\\
8.0e+06& 3.6217e+24& 1.4503e+06& 3.6711e-13& 1.7574e+12& 6.0245e+05& 2.7611e+33\\
9.5e+06& 4.3008e+24& 9.9237e+05& 5.3651e-13& 1.2025e+12& 4.1224e+05& 4.0351e+33\\
\cutinhead{$L_{xx}=$ 2.250e+49, $A_r=$ 1.0667, $\chi _\mathrm{now}=$ 78.0010, $F_{r\left[\mathrm{now}\right]}=$ 0.2080}
5.0e+05& 2.2842e+23& 2.4721e+08& 1.9654e-15& 3.2365e+14& 1.4379e+08& 1.8697e+31\\
2.0e+06& 9.1370e+23& 1.6032e+07& 3.0306e-14& 2.0990e+13& 9.3251e+06& 2.8830e+32\\
3.5e+06& 1.5990e+24& 5.2331e+06& 9.2847e-14& 6.8512e+12& 3.0438e+06& 8.8325e+32\\
5.0e+06& 2.2842e+24& 2.5674e+06& 1.8925e-13& 3.3612e+12& 1.4933e+06& 1.8003e+33\\
6.5e+06& 2.9695e+24& 1.4665e+06& 3.3132e-13& 1.9200e+12& 8.5297e+05& 3.1518e+33\\
8.0e+06& 3.6548e+24& 1.0018e+06& 4.8500e-13& 1.3116e+12& 5.8269e+05& 4.6138e+33\\
9.5e+06& 4.3401e+24& 6.8537e+05& 7.0892e-13& 8.9730e+11& 3.9864e+05& 6.7439e+33\\
\cutinhead{$L_{xx}=$ 2.450e+49, $A_r=$ 1.0781, $\chi _\mathrm{now}=$ 85.1031, $F_{r\left[\mathrm{now}\right]}=$ 0.0854}
5.0e+05& 2.3066e+23& 1.6364e+08& 2.7057e-15& 2.2817e+14& 1.2169e+08& 3.1806e+31\\
2.0e+06& 9.2266e+23& 1.0612e+07& 4.1721e-14& 1.4798e+13& 7.8918e+06& 4.9043e+32\\
3.5e+06& 1.6146e+24& 3.4643e+06& 1.2780e-13& 4.8306e+12& 2.5762e+06& 1.5024e+33\\
5.0e+06& 2.3066e+24& 1.6998e+06& 2.6048e-13& 2.3702e+12& 1.2640e+06& 3.0619e+33\\
6.5e+06& 2.9986e+24& 9.7051e+05& 4.5621e-13& 1.3533e+12& 7.2171e+05& 5.3628e+33\\
8.0e+06& 3.6906e+24& 6.6332e+05& 6.6749e-13& 9.2492e+11& 4.9328e+05& 7.8463e+33\\
9.5e+06& 4.3826e+24& 4.5373e+05& 9.7583e-13& 6.3266e+11& 3.3741e+05& 1.1471e+34\\
\cutinhead{$L_{xx}=$ 2.515e+49, $A_r=$ 1.0821, $\chi _\mathrm{now}=$ 86.6041, $F_{r\left[\mathrm{now}\right]}=$ 0.0593}
5.0e+05& 2.3143e+23& 1.4302e+08& 3.0116e-15& 2.0335e+14& 1.1392e+08& 3.7690e+31\\
2.0e+06& 9.2571e+23& 9.2751e+06& 4.6439e-14& 1.3187e+13& 7.3879e+06& 5.8119e+32\\
3.5e+06& 1.6200e+24& 3.0277e+06& 1.4226e-13& 4.3048e+12& 2.4117e+06& 1.7804e+33\\
5.0e+06& 2.3143e+24& 1.4855e+06& 2.8996e-13& 2.1121e+12& 1.1832e+06& 3.6288e+33\\
6.5e+06& 3.0086e+24& 8.4850e+05& 5.0763e-13& 1.2064e+12& 6.7586e+05& 6.3530e+33\\
8.0e+06& 3.7028e+24& 5.7974e+05& 7.4297e-13& 8.2427e+11& 4.6178e+05& 9.2982e+33\\
9.5e+06& 4.3971e+24& 3.9642e+05& 1.0865e-12& 5.6363e+11& 3.1576e+05& 1.3598e+34\\
\enddata
\end{deluxetable}

\begin{deluxetable}{ccccccccc}
\tablewidth{0pt}
\tablecaption{Case $M/M_\odot =1.32$:~Results concerning the aligned rotator (part I)\label{tab:132_Aligned_constant_for_Bs}}
\tablehead{
\colhead{$L_{xx}$}& \colhead{$P_{xx}$}& \colhead{$T_{xx}$}& \colhead{$I_{11}$}& \colhead{$I_{33}$}& \colhead{$\left\langle H_{ts}\right\rangle$}& \colhead{$H_{t \left[\mathrm{max}\right]}$}& \colhead{$\left\langle H_t\right\rangle$}& \colhead{$\omega_\mathrm{e}^{-1}$}\\
}
\tabletypesize{\scriptsize}
\startdata
1.450e+49& 7.9869& 3.4198e+48& 3.4681e+49& 3.4068e+49& 1.1345e+09& 7.6962e+12& 2.6052e+12& 5.5989\\
1.500e+49& 7.7621& 3.6403e+48& 3.4844e+49& 3.4246e+49& 1.1366e+09& 7.6564e+12& 2.5930e+12& 5.5919\\
1.800e+49& 6.7139& 5.0661e+48& 3.5905e+49& 3.5413e+49& 1.1505e+09& 7.4057e+12& 2.5160e+12& 5.5593\\
2.100e+49& 5.9828& 6.6382e+48& 3.7117e+49& 3.6756e+49& 1.1660e+09& 7.1377e+12& 2.4335e+12& 5.5200\\
2.400e+49& 5.4567& 8.3170e+48& 3.8489e+49& 3.8283e+49& 1.1831e+09& 6.8549e+12& 2.3461e+12& 5.4845\\
2.700e+49& 5.0875& 1.0073e+49& 4.0003e+49& 3.9979e+49& 1.2016e+09& 6.5663e+12& 2.2566e+12& 5.4371\\
2.730e+49& 5.0569& 1.0252e+49& 4.0161e+49& 4.0158e+49& 1.2035e+09& 6.5374e+12& 2.2476e+12& 5.4312\\
\enddata
\end{deluxetable}

\begin{deluxetable}{cccccc}
\tablewidth{0pt}
\tablecaption{Case $M/M_\odot =1.32$:~Results concerning the aligned rotator (part II)\label{tab:132_Aligned_variable_for_Bs}}
\tablehead{
\colhead{$B_s$}& \colhead{$\beta _*^p$}& \colhead{$V_{As}$}& \colhead{$t_{As}$}& \colhead{$H_{p \left[\mathrm{max}\right]}$}& \colhead{$\left\langle H_p\right\rangle$}\\
}
\tabletypesize{\scriptsize}
\startdata
\cutinhead{Models with $L_{xx}=$ 1.450e+49}
5.0e+05& 1.1063e+05& 1.4001e+03& 2.1902e+05& 2.2045e+08& 5.2208e+07\\
2.0e+06& 2.7657e+04& 5.6004e+03& 5.4756e+04& 8.8179e+08& 2.0883e+08\\
3.5e+06& 1.5803e+04& 9.8013e+03& 3.1287e+04& 1.5432e+09& 3.6548e+08\\
5.0e+06& 1.1062e+04& 1.4001e+04& 2.1902e+04& 2.2045e+09& 5.2209e+08\\
6.5e+06& 8.5105e+03& 1.8200e+04& 1.6849e+04& 2.8656e+09& 6.7864e+08\\
8.0e+06& 6.9142e+03& 2.2401e+04& 1.3689e+04& 3.5271e+09& 8.3532e+08\\
9.5e+06& 5.8227e+03& 2.6601e+04& 1.1528e+04& 4.1883e+09& 9.9191e+08\\
\cutinhead{Models with $L_{xx}=$ 1.500e+49}
5.0e+05& 1.1064e+05& 1.4001e+03& 2.1944e+05& 2.1926e+08& 5.1959e+07\\
2.0e+06& 2.7660e+04& 5.6006e+03& 5.4858e+04& 8.7709e+08& 2.0784e+08\\
3.5e+06& 1.5806e+04& 9.8007e+03& 3.1349e+04& 1.5348e+09& 3.6371e+08\\
5.0e+06& 1.1064e+04& 1.4001e+04& 2.1943e+04& 2.1927e+09& 5.1960e+08\\
6.5e+06& 8.5105e+03& 1.8203e+04& 1.6879e+04& 2.8507e+09& 6.7552e+08\\
8.0e+06& 6.9161e+03& 2.2399e+04& 1.3716e+04& 3.5079e+09& 8.3125e+08\\
9.5e+06& 5.8227e+03& 2.6605e+04& 1.1548e+04& 4.1665e+09& 9.8734e+08\\
\cutinhead{Models with $L_{xx}=$ 1.800e+49}
5.0e+05& 1.1076e+05& 1.4001e+03& 2.2214e+05& 2.1182e+08& 5.0387e+07\\
2.0e+06& 2.7690e+04& 5.6006e+03& 5.5532e+04& 8.4729e+08& 2.0155e+08\\
3.5e+06& 1.5823e+04& 9.8009e+03& 3.1733e+04& 1.4827e+09& 3.5271e+08\\
5.0e+06& 1.1077e+04& 1.4000e+04& 2.2215e+04& 2.1180e+09& 5.0383e+08\\
6.5e+06& 8.5198e+03& 1.8202e+04& 1.7086e+04& 2.7538e+09& 6.5506e+08\\
8.0e+06& 6.9235e+03& 2.2399e+04& 1.3885e+04& 3.3887e+09& 8.0610e+08\\
9.5e+06& 5.8301e+03& 2.6600e+04& 1.1692e+04& 4.0242e+09& 9.5727e+08\\
\cutinhead{Models with $L_{xx}=$ 2.100e+49}
5.0e+05& 1.1089e+05& 1.4001e+03& 2.2515e+05& 2.0388e+08& 4.8705e+07\\
2.0e+06& 2.7722e+04& 5.6006e+03& 5.6286e+04& 8.1553e+08& 1.9483e+08\\
3.5e+06& 1.5842e+04& 9.8005e+03& 3.2165e+04& 1.4271e+09& 3.4093e+08\\
5.0e+06& 1.1088e+04& 1.4002e+04& 2.2514e+04& 2.0389e+09& 4.8708e+08\\
6.5e+06& 8.5291e+03& 1.8203e+04& 1.7317e+04& 2.6507e+09& 6.3323e+08\\
8.0e+06& 6.9310e+03& 2.2401e+04& 1.4072e+04& 3.2619e+09& 7.7925e+08\\
9.5e+06& 5.8357e+03& 2.6605e+04& 1.1849e+04& 3.8741e+09& 9.2549e+08\\
\cutinhead{Models with $L_{xx}=$ 2.400e+49}
5.0e+05& 1.1102e+05& 1.4001e+03& 2.2849e+05& 1.9553e+08& 4.6931e+07\\
2.0e+06& 2.7755e+04& 5.6003e+03& 5.7124e+04& 7.8209e+08& 1.8772e+08\\
3.5e+06& 1.5860e+04& 9.8004e+03& 3.2643e+04& 1.3686e+09& 3.2850e+08\\
5.0e+06& 1.1101e+04& 1.4002e+04& 2.2848e+04& 1.9553e+09& 4.6933e+08\\
6.5e+06& 8.5403e+03& 1.8201e+04& 1.7577e+04& 2.5417e+09& 6.1007e+08\\
8.0e+06& 6.9384e+03& 2.2403e+04& 1.4280e+04& 3.1285e+09& 7.5092e+08\\
9.5e+06& 5.8432e+03& 2.6602e+04& 1.2026e+04& 3.7149e+09& 8.9167e+08\\
\cutinhead{Models with $L_{xx}=$ 2.700e+49}
5.0e+05& 1.1115e+05& 1.4001e+03& 2.3209e+05& 1.8703e+08& 4.5117e+07\\
2.0e+06& 2.7787e+04& 5.6005e+03& 5.8022e+04& 7.4811e+08& 1.8047e+08\\
3.5e+06& 1.5879e+04& 9.8003e+03& 3.3157e+04& 1.3091e+09& 3.1581e+08\\
5.0e+06& 1.1114e+04& 1.4002e+04& 2.3208e+04& 1.8703e+09& 4.5119e+08\\
6.5e+06& 8.5496e+03& 1.8202e+04& 1.7852e+04& 2.4314e+09& 5.8654e+08\\
8.0e+06& 6.9459e+03& 2.2405e+04& 1.4504e+04& 2.9928e+09& 7.2197e+08\\
9.5e+06& 5.8506e+03& 2.6599e+04& 1.2217e+04& 3.5531e+09& 8.5712e+08\\
\cutinhead{Models with $L_{xx}=$ 2.730e+49}
5.0e+05& 1.1116e+05& 1.4001e+03& 2.3246e+05& 1.8618e+08& 4.4936e+07\\
2.0e+06& 2.7791e+04& 5.6004e+03& 5.8116e+04& 7.4469e+08& 1.7974e+08\\
3.5e+06& 1.5881e+04& 9.8003e+03& 3.3210e+04& 1.3032e+09& 3.1453e+08\\
5.0e+06& 1.1116e+04& 1.4001e+04& 2.3246e+04& 1.8617e+09& 4.4935e+08\\
6.5e+06& 8.5515e+03& 1.8200e+04& 1.7883e+04& 2.4201e+09& 5.8412e+08\\
8.0e+06& 6.9477e+03& 2.2401e+04& 1.4529e+04& 2.9787e+09& 7.1895e+08\\
9.5e+06& 5.8506e+03& 2.6602e+04& 1.2235e+04& 3.5373e+09& 8.5377e+08\\
\enddata
\end{deluxetable}

\begin{deluxetable}{ccccccccc}
\tablewidth{0pt}
\tablecaption{Case $M/M_\odot =1.32$:~Results concerning the perpendicular rotator (part I)\label{tab:132_Perpendicular_constant_for_Bs}}
\tablehead{
\colhead{$L_{xx}$}& \colhead{$P_\mathrm{RR}$}& \colhead{$T_\mathrm{RR}$}& \colhead{$I_{11}$}& \colhead{$I_{33}$}& \colhead{$\left\langle H_{ts}\right\rangle$}& \colhead{$H_{t \left[\mathrm{max}\right]}$}& \colhead{$\left\langle H_t\right\rangle$}& \colhead{$\omega_\mathrm{e}^{-1}$}\\
}
\tabletypesize{\scriptsize}
\startdata
1.450e+49& 14.8896& 3.0594e+48& 3.4362e+49& 3.1499e+49& 1.1025e+09& 8.3213e+12& 2.7960e+12& 1.0000\\
1.500e+49& 14.4496& 3.2613e+48& 3.4496e+49& 3.1499e+49& 1.1025e+09& 8.3213e+12& 2.7960e+12& 1.0000\\
1.800e+49& 12.3443& 4.5810e+48& 3.5364e+49& 3.1499e+49& 1.1025e+09& 8.3213e+12& 2.7960e+12& 1.0000\\
2.100e+49& 10.8524& 6.0792e+48& 3.6271e+49& 3.1499e+49& 1.1025e+09& 8.3213e+12& 2.7960e+12& 1.0000\\
2.400e+49& \phn 9.7603& 7.7250e+48& 3.7282e+49& 3.1499e+49& 1.1025e+09& 8.3213e+12& 2.7960e+12& 1.0000\\
2.700e+49& \phn 8.9148& 9.5149e+48& 3.8309e+49& 3.1499e+49& 1.1025e+09& 8.3213e+12& 2.7960e+12& 1.0000\\
2.730e+49& \phn 8.8410& 9.7009e+48& 3.8413e+49& 3.1499e+49& 1.1025e+09& 8.3213e+12& 2.7960e+12& 1.0000\\
\enddata
\end{deluxetable}

\begin{deluxetable}{cccccc}
\tablewidth{0pt}
\tablecaption{Case $M/M_\odot =1.32$:~Results concerning the perpendicular rotator (part II)\label{tab:132_Perpendicular_variable_for_Bs}}
\tablehead{
\colhead{$B_s$}& \colhead{$\beta _*^p$}& \colhead{$V_{As}$}& \colhead{$t_{As}$}& \colhead{$H_{p \left[\mathrm{max}\right]}$}& \colhead{$\left\langle H_p\right\rangle$}\\
}
\tabletypesize{\scriptsize}
\startdata
\cutinhead{Models with $L_{xx}=$ 1.450e+49}
5.3e+05& 1.0358e+05& 1.4914e+03& 1.9978e+05& 2.5468e+08& 5.9781e+07\\
2.1e+06& 2.5894e+04& 5.9654e+03& 4.9945e+04& 1.0187e+09& 2.3912e+08\\
3.7e+06& 1.4797e+04& 1.0439e+04& 2.8541e+04& 1.7827e+09& 4.1844e+08\\
5.3e+06& 1.0357e+04& 1.4915e+04& 1.9977e+04& 2.5470e+09& 5.9784e+08\\
6.9e+06& 7.9670e+03& 1.9389e+04& 1.5367e+04& 3.3110e+09& 7.7718e+08\\
8.5e+06& 6.4736e+03& 2.3862e+04& 1.2486e+04& 4.0748e+09& 9.5648e+08\\
1.0e+07& 5.4512e+03& 2.8337e+04& 1.0515e+04& 4.8390e+09& 1.1359e+09\\
\cutinhead{Models with $L_{xx}=$ 1.500e+49}
5.3e+05& 1.0314e+05& 1.4976e+03& 1.9895e+05& 2.5575e+08& 6.0031e+07\\
2.1e+06& 2.5785e+04& 5.9906e+03& 4.9736e+04& 1.0230e+09& 2.4013e+08\\
3.7e+06& 1.4734e+04& 1.0484e+04& 2.8420e+04& 1.7903e+09& 4.2023e+08\\
5.3e+06& 1.0315e+04& 1.4975e+04& 1.9896e+04& 2.5573e+09& 6.0028e+08\\
7.0e+06& 7.9346e+03& 1.9468e+04& 1.5304e+04& 3.3245e+09& 7.8036e+08\\
8.6e+06& 6.4469e+03& 2.3960e+04& 1.2435e+04& 4.0917e+09& 9.6044e+08\\
1.0e+07& 5.4284e+03& 2.8456e+04& 1.0470e+04& 4.8594e+09& 1.1406e+09\\
\cutinhead{Models with $L_{xx}=$ 1.800e+49}
5.5e+05& 1.0041e+05& 1.5384e+03& 1.9367e+05& 2.6271e+08& 6.1666e+07\\
2.2e+06& 2.5103e+04& 6.1535e+03& 4.8419e+04& 1.0508e+09& 2.4666e+08\\
3.8e+06& 1.4343e+04& 1.0770e+04& 2.7666e+04& 1.8391e+09& 4.3169e+08\\
5.5e+06& 1.0040e+04& 1.5385e+04& 1.9366e+04& 2.6273e+09& 6.1670e+08\\
7.1e+06& 7.7229e+03& 2.0002e+04& 1.4896e+04& 3.4157e+09& 8.0175e+08\\
8.8e+06& 6.2752e+03& 2.4616e+04& 1.2104e+04& 4.2036e+09& 9.8671e+08\\
1.0e+07& 5.2853e+03& 2.9226e+04& 1.0194e+04& 4.9910e+09& 1.1715e+09\\
\cutinhead{Models with $L_{xx}=$ 2.100e+49}
5.7e+05& 9.7465e+04& 1.5849e+03& 1.8799e+05& 2.7065e+08& 6.3528e+07\\
2.3e+06& 2.4366e+04& 6.3395e+03& 4.6999e+04& 1.0826e+09& 2.5411e+08\\
4.0e+06& 1.3924e+04& 1.1094e+04& 2.6856e+04& 1.8945e+09& 4.4470e+08\\
5.7e+06& 9.7466e+03& 1.5849e+04& 1.8799e+04& 2.7065e+09& 6.3528e+08\\
7.4e+06& 7.4978e+03& 2.0602e+04& 1.4462e+04& 3.5182e+09& 8.2582e+08\\
9.1e+06& 6.0921e+03& 2.5356e+04& 1.1751e+04& 4.3300e+09& 1.0164e+09\\
1.1e+07& 5.1289e+03& 3.0118e+04& 9.8928e+03& 5.1432e+09& 1.2072e+09\\
\cutinhead{Models with $L_{xx}=$ 2.400e+49}
5.8e+05& 9.4339e+04& 1.6374e+03& 1.8196e+05& 2.7962e+08& 6.5634e+07\\
2.3e+06& 2.3584e+04& 6.5497e+03& 4.5490e+04& 1.1185e+09& 2.6254e+08\\
4.1e+06& 1.3477e+04& 1.1461e+04& 2.5996e+04& 1.9573e+09& 4.5942e+08\\
5.8e+06& 9.4338e+03& 1.6374e+04& 1.8196e+04& 2.7962e+09& 6.5635e+08\\
7.6e+06& 7.2575e+03& 2.1284e+04& 1.3998e+04& 3.6347e+09& 8.5316e+08\\
9.4e+06& 5.8957e+03& 2.6201e+04& 1.1372e+04& 4.4743e+09& 1.0502e+09\\
1.1e+07& 4.9649e+03& 3.1113e+04& 9.5764e+03& 5.3131e+09& 1.2471e+09\\
\cutinhead{Models with $L_{xx}=$ 2.700e+49}
6.1e+05& 9.1131e+04& 1.6950e+03& 1.7578e+05& 2.8946e+08& 6.7944e+07\\
2.4e+06& 2.2783e+04& 6.7800e+03& 4.3945e+04& 1.1578e+09& 2.7177e+08\\
4.2e+06& 1.3020e+04& 1.1864e+04& 2.5113e+04& 2.0261e+09& 4.7558e+08\\
6.1e+06& 9.1133e+03& 1.6950e+04& 1.7578e+04& 2.8945e+09& 6.7942e+08\\
7.9e+06& 7.0095e+03& 2.2037e+04& 1.3520e+04& 3.7633e+09& 8.8334e+08\\
9.7e+06& 5.6954e+03& 2.7122e+04& 1.0985e+04& 4.6316e+09& 1.0872e+09\\
1.1e+07& 4.7970e+03& 3.2201e+04& 9.2526e+03& 5.4990e+09& 1.2908e+09\\
\cutinhead{Models with $L_{xx}=$ 2.730e+49}
6.1e+05& 9.0809e+04& 1.7010e+03& 1.7515e+05& 2.9049e+08& 6.8185e+07\\
2.4e+06& 2.2701e+04& 6.8045e+03& 4.3787e+04& 1.1620e+09& 2.7275e+08\\
4.3e+06& 1.2972e+04& 1.1908e+04& 2.5021e+04& 2.0335e+09& 4.7733e+08\\
6.1e+06& 9.0809e+03& 1.7010e+04& 1.7516e+04& 2.9049e+09& 6.8185e+08\\
7.9e+06& 6.9848e+03& 2.2115e+04& 1.3472e+04& 3.7766e+09& 8.8648e+08\\
9.7e+06& 5.6763e+03& 2.7213e+04& 1.0949e+04& 4.6472e+09& 1.0908e+09\\
1.2e+07& 4.7799e+03& 3.2317e+04& 9.2195e+03& 5.5187e+09& 1.2954e+09\\
\enddata
\end{deluxetable}

\begin{deluxetable}{ccccccccccc}
\tablewidth{0pt}
\tablecaption{Case $M/M_\odot =1.32$:~Results concerning the turn-over phase\label{tab:132_Turn_Over}}
\tablehead{
\colhead{$B_s$}& \colhead{$f$}& \colhead{$t_\mathrm{TOV}$}& \colhead{$\overset{.}{P}_\mathrm{TOV}$}& \colhead{$N_{xx}$}& \colhead{$t_\mathrm{now}$}& \colhead{$\overset{.}{T}$}\\
}
\tabletypesize{\scriptsize}
\startdata
\cutinhead{$L_{xx}=$ 1.450e+49, $A_r=$ 1.0630, $\chi _\mathrm{now}=$ 2.6088, $F_{r\left[\mathrm{now}\right]}=$ 0.9993}
5.0e+05& 5.0033e+22& 5.1102e+09& 4.2833e-17& 2.0178e+16& 9.7247e+06& 2.2362e+30\\
2.0e+06& 2.0013e+23& 2.9030e+08& 7.5400e-16& 1.1462e+15& 5.5243e+05& 3.9365e+31\\
3.5e+06& 3.5023e+23& 1.0109e+08& 2.1652e-15& 3.9917e+14& 1.9238e+05& 1.1304e+32\\
5.0e+06& 5.0033e+23& 5.1160e+07& 4.2785e-15& 2.0200e+14& 9.7356e+04& 2.2337e+32\\
6.5e+06& 6.5043e+23& 3.0327e+07& 7.2174e-15& 1.1975e+14& 5.7712e+04& 3.7681e+32\\
8.0e+06& 8.0053e+23& 1.9320e+07& 1.1329e-14& 7.6285e+13& 3.6766e+04& 5.9149e+32\\
9.5e+06& 9.5063e+23& 1.3724e+07& 1.5949e-14& 5.4189e+13& 2.6117e+04& 8.3267e+32\\
\cutinhead{$L_{xx}=$ 1.500e+49, $A_r=$ 1.0672, $\chi _\mathrm{now}=$ 10.5276, $F_{r\left[\mathrm{now}\right]}=$ 0.9835}
5.0e+05& 5.0243e+22& 4.4365e+09& 4.7799e-17& 1.8025e+16& 1.5783e+08& 2.7093e+30\\
2.0e+06& 2.0097e+23& 2.5201e+08& 8.4148e-16& 1.0239e+15& 8.9652e+06& 4.7696e+31\\
3.5e+06& 3.5170e+23& 8.7777e+07& 2.4159e-15& 3.5662e+14& 3.1226e+06& 1.3694e+32\\
5.0e+06& 5.0243e+23& 4.4414e+07& 4.7746e-15& 1.8045e+14& 1.5800e+06& 2.7063e+32\\
6.5e+06& 6.5316e+23& 2.6320e+07& 8.0571e-15& 1.0693e+14& 9.3632e+05& 4.5669e+32\\
8.0e+06& 8.0388e+23& 1.6776e+07& 1.2641e-14& 6.8157e+13& 5.9680e+05& 7.1650e+32\\
9.5e+06& 9.5461e+23& 1.1911e+07& 1.7804e-14& 4.8390e+13& 4.2372e+05& 1.0092e+33\\
\cutinhead{$L_{xx}=$ 1.800e+49, $A_r=$ 1.0940, $\chi _\mathrm{now}=$ 45.5260, $F_{r\left[\mathrm{now}\right]}=$ 0.7008}
5.0e+05& 5.1611e+22& 2.0819e+09& 8.5756e-17& 9.7791e+15& 4.7557e+08& 7.3889e+30\\
2.0e+06& 2.0645e+23& 1.1827e+08& 1.5096e-15& 5.5552e+14& 2.7015e+07& 1.3007e+32\\
3.5e+06& 3.6128e+23& 4.1189e+07& 4.3346e-15& 1.9347e+14& 9.4086e+06& 3.7348e+32\\
5.0e+06& 5.1611e+23& 2.0846e+07& 8.5646e-15& 9.7917e+13& 4.7618e+06& 7.3794e+32\\
6.5e+06& 6.7095e+23& 1.2351e+07& 1.4455e-14& 5.8014e+13& 2.8213e+06& 1.2455e+33\\
8.0e+06& 8.2578e+23& 7.8719e+06& 2.2681e-14& 3.6975e+13& 1.7981e+06& 1.9542e+33\\
9.5e+06& 9.8062e+23& 5.5914e+06& 3.1931e-14& 2.6264e+13& 1.2772e+06& 2.7512e+33\\
\cutinhead{$L_{xx}=$ 2.100e+49, $A_r=$ 1.1221, $\chi _\mathrm{now}=$ 78.1579, $F_{r\left[\mathrm{now}\right]}=$ 0.2053}
5.0e+05& 5.3169e+22& 1.0344e+09& 1.4928e-16& 5.4526e+15& 4.2851e+08& 1.7137e+31\\
2.0e+06& 2.1268e+23& 5.8767e+07& 2.6276e-15& 3.0977e+14& 2.4344e+07& 3.0165e+32\\
3.5e+06& 3.7219e+23& 2.0466e+07& 7.5448e-15& 1.0788e+14& 8.4782e+06& 8.6615e+32\\
5.0e+06& 5.3169e+23& 1.0355e+07& 1.4913e-14& 5.4581e+13& 4.2894e+06& 1.7120e+33\\
6.5e+06& 6.9120e+23& 6.1359e+06& 2.5166e-14& 3.2343e+13& 2.5418e+06& 2.8891e+33\\
8.0e+06& 8.5071e+23& 3.9104e+06& 3.9488e-14& 2.0612e+13& 1.6199e+06& 4.5333e+33\\
9.5e+06& 1.0102e+24& 2.7771e+06& 5.5603e-14& 1.4638e+13& 1.1504e+06& 6.3833e+33\\
\cutinhead{$L_{xx}=$ 2.400e+49, $A_r=$ 1.1534, $\chi _\mathrm{now}=$ 88.1526, $F_{r\left[\mathrm{now}\right]}=$ 0.0322}
5.0e+05& 5.4929e+22& 5.4496e+08& 2.5042e-16& 3.1495e+15& 3.2205e+08& 3.4448e+31\\
2.0e+06& 2.1972e+23& 3.0965e+07& 4.4072e-15& 1.7896e+14& 1.8300e+07& 6.0626e+32\\
3.5e+06& 3.8450e+23& 1.0782e+07& 1.2657e-14& 6.2315e+13& 6.3721e+06& 1.7411e+33\\
5.0e+06& 5.4929e+23& 5.4551e+06& 2.5016e-14& 3.1527e+13& 3.2238e+06& 3.4413e+33\\
6.5e+06& 7.1407e+23& 3.2334e+06& 4.2207e-14& 1.8687e+13& 1.9108e+06& 5.8060e+33\\
8.0e+06& 8.7886e+23& 2.0596e+06& 6.6260e-14& 1.1903e+13& 1.2172e+06& 9.1148e+33\\
9.5e+06& 1.0436e+24& 1.4634e+06& 9.3258e-14& 8.4572e+12& 8.6480e+05& 1.2829e+34\\
\cutinhead{$L_{xx}=$ 2.700e+49, $A_r=$ 1.1851, $\chi _\mathrm{now}=$ 89.7462, $F_{r\left[\mathrm{now}\right]}=$ 0.0044}
5.0e+05& 5.6865e+22& 2.8011e+08& 4.3328e-16& 1.7363e+15& 2.1316e+08& 6.3199e+31\\
2.0e+06& 2.2746e+23& 1.5521e+07& 7.8191e-15& 9.6214e+13& 1.1812e+07& 1.1405e+33\\
3.5e+06& 3.9806e+23& 5.5422e+06& 2.1898e-14& 3.4355e+13& 4.2175e+06& 3.1942e+33\\
5.0e+06& 5.6865e+23& 2.8039e+06& 4.3284e-14& 1.7381e+13& 2.1338e+06& 6.3135e+33\\
6.5e+06& 7.3925e+23& 1.6616e+06& 7.3040e-14& 1.0300e+13& 1.2645e+06& 1.0654e+34\\
8.0e+06& 9.0984e+23& 1.0584e+06& 1.1467e-13& 6.5605e+12& 8.0539e+05& 1.6727e+34\\
9.5e+06& 1.0804e+24& 7.5232e+05& 1.6132e-13& 4.6635e+12& 5.7250e+05& 2.3531e+34\\
\cutinhead{$L_{xx}=$ 2.730e+49, $A_r=$ 1.1884, $\chi _\mathrm{now}=$ 89.7936, $F_{r\left[\mathrm{now}\right]}=$ 0.0036}
5.0e+05& 5.7068e+22& 2.6103e+08& 4.5968e-16& 1.6279e+15& 2.0302e+08& 6.6925e+31\\
2.0e+06& 2.2827e+23& 1.4465e+07& 8.2952e-15& 9.0209e+13& 1.1251e+07& 1.2077e+33\\
3.5e+06& 3.9947e+23& 5.1648e+06& 2.3233e-14& 3.2209e+13& 4.0170e+06& 3.3824e+33\\
5.0e+06& 5.7068e+23& 2.6133e+06& 4.5916e-14& 1.6297e+13& 2.0325e+06& 6.6849e+33\\
6.5e+06& 7.4188e+23& 1.5488e+06& 7.7475e-14& 9.6587e+12& 1.2046e+06& 1.1279e+34\\
8.0e+06& 9.1309e+23& 9.8659e+05& 1.2162e-13& 6.1526e+12& 7.6732e+05& 1.7707e+34\\
9.5e+06& 1.0843e+24& 7.0094e+05& 1.7119e-13& 4.3712e+12& 5.4516e+05& 2.4923e+34\\
\enddata
\end{deluxetable}

\end{document}